\documentclass[twocolumn,english,aps,pre]{revtex4}
\usepackage[T1]{fontenc}
\usepackage[latin9]{inputenc}
\usepackage{color}
\usepackage{babel}
\usepackage{amsmath}
\usepackage{graphicx}
\usepackage[bookmarks=false,
 breaklinks=false,pdfborder={0 0 1},backref=section,colorlinks=false]
 {hyperref}
\hypersetup{
 colorlinks,linkcolor=blue,urlcolor=black,citecolor=blue,anchorcolor=blue}
\makeatletter

\makeatletter

\makeatletter
\@ifundefined{textcolor}{}
{%
 \definecolor{BLACK}{gray}{0}
 \definecolor{WHITE}{gray}{1}
 \definecolor{RED}{rgb}{1,0,0}
 \definecolor{GREEN}{rgb}{0,1,0}
 \definecolor{BLUE}{rgb}{0,0,1}
 \definecolor{CYAN}{cmyk}{1,0,0,0}
 \definecolor{MAGENTA}{cmyk}{0,1,0,0}
 \definecolor{YELLOW}{cmyk}{0,0,1,0}
}

\makeatother

\begin{document}
\title{Conformable scaling and critical phenomena: A unified framework for
phase transitions}
\author{Jos{\'e} Weberszpil}
\email{josewebe@gmail.com}

\affiliation{Universidade Federal Rural do Rio de Janeiro, UFRRJ-DEFIS/ICE; BR-465,
Km 7 Serop{\'e}dica-Rio de Janeiro CEP: 23.897-000}
\author{Ralf Metzler}
\email{ralf.metzler@uni-potsdam.de}

\affiliation{University of Potsdam, Institute of Physics \& Astronomy, Karl-Liebknecht-St.
24/25, 14476 Potsdam-Golm, Germany}
\date{\today}
\begin{abstract}
We investigate the application of conformable derivatives to model
critical phenomena near continuous phase transition points. By incorporating
a deformation parameter into the differential structure, we derive
unified expressions for thermodynamic observables such as heat capacity,
magnetization, susceptibility, and coherence length, each exhibiting
a power-law behavior near the critical temperature. The conformable
derivative framework naturally embeds scale invariance and critical
slowing down into the dynamics without resorting to fully nonlocal
fractional calculus. Modified Ginzburg--Landau equations are constructed
to model superconducting transitions, leading to analytical expressions
for the order parameter and the London penetration depth. Experimental
data from niobium confirm the model's applicability, showing excellent
fits and capturing asymmetric scaling behavior around the critical
point. This work offers a bridge between classical mean-field theory
and generalized scaling frameworks, with implications for both theoretical
modeling and experimental analysis. 
\end{abstract}
\maketitle
\vspace{0.5cm}

\section{Introduction}

Critical phenomena play a central role in the understanding of phase
transitions, where physical observables such as the heat capacity,
magnetization, and susceptibility exhibit a non-analytical behavior
characterized by power-law divergences near a critical temperature
$T_{c}$ \cite{Landau1978StatisticalPart1,goldenfeld1992,stanley1971}.
These behaviors are traditionally described through scaling hypotheses
and renormalization group (RG) analysis, which explain the emergence
of universality across diverse systems. These phenomena range from
magnetic materials and superconductors to liquid-gas transitions and
even biological systems \cite{goldenfeld1992,stanley1971}. Near
critical points, physical systems exhibit remarkable scaling behavior
characterized by power-law divergences of thermodynamic quantities,
with critical exponents that depend only on fundamental symmetries
and spatial dimensionality rather than microscopic details \cite{Wilson1971,Fisher1972}.
This universality principle, established through decades of theoretical
and experimental work, forms the cornerstone of the modern theory
of critical phenomena.

The mathematical description of critical behavior has evolved significantly
since the early phenomenological approaches of Landau and Ginzburg
\cite{Landau1937Eng,Ginzburg1950Eng}. The subsequent development
of Ginzburg-Landau theory, particularly as systematized by de Gennes
\cite{deGennes1966}, provided a robust framework for understanding
superconducting phase transitions and scaling behavior near critical
points. The development of renormalization group theory by Wilson
and others \cite{Wilson1971,Wilson1974} provided deep insights into
the origin of universality and scaling laws, while field-theoretic
methods enabled precise calculations of critical exponents \cite{ZinnJustin2021}.
These advances established that critical phenomena emerge from the
competition between thermal fluctuations and ordering tendencies,
with correlation lengths diverging as $\xi\sim|T-T_{c}|^{-\nu}$ and
other observables following characteristic power-laws.

Despite these theoretical successes, many experimental systems exhibit
deviations from ideal critical behavior due to finite-size effects,
quenched disorder, non-equilibrium conditions, and other realistic
complications \cite{Privman1990,Harris1974,Hohenberg1977}. There
remains a need for alternative frameworks that provide analytical
tractability, flexible modeling, and connections to generalized thermodynamic
formalisms. This is particularly relevant in systems where microscopic
details are poorly known, long-range interactions are present, or
nonlocal and memory effects become significant. Real materials often
deviate from ideal critical behavior due to quenched disorder, finite-size
effects, and non-equilibrium conditions. These deviations often manifest
as modified scaling laws, anomalous relaxation dynamics, and effective
critical exponents that differ from theoretical predictions. Understanding
and modeling such complex critical behavior remains an active area
of research with important implications for materials science, statistical
physics, and complex systems.

In this work, we explore a deformation-based approach to critical
phenomena using \emph{conformable derivatives} \cite{Weberszpil2015,weberszpil2016variational,Weberszpil2017_Tissue,weberszpil2017generalized,Sotolongo2021,rosa2018dual,godinho2020variational,Khalil2014},
a form of generalized calculus that introduces a deformation parameter
$\mu$ into the derivative operator: 
\begin{equation}
D_{T}^{(\mu)}f(T):=T^{1-\mu}\frac{df}{dT}.\label{eq:Conformable Derivative}
\end{equation}
This operator naturally yields temperature-dependent scaling behavior
and introduces power-law features without requiring full fractional
integration or nonlocal convolution terms. It thus offers an intermediate
practical framework settled between classical and fractional derivatives;
and retains analytical simplicity while capturing essential scaling
features.

The conformable derivative framework addresses specific limitations
in existing approaches to critical phenomena. Real materials exhibit:
(1) \emph{Spatial inhomogeneities\/} leading to effective fractal
geometries with non-integer dimensions; (2) \emph{Temporal memory
effects\/} from quenched disorder or slow relaxation processes; and
(3) \emph{Finite-size constraints\/} that modify critical scaling.
Each of these physical mechanisms naturally leads to temperature-dependent
weighting factors of the form $T^{1-\mu}$, providing direct physical
justification for definition (\ref{eq:Conformable Derivative}). Unlike
purely phenomenological approaches, the conformable-derivative framework
connects the deformation parameter $\mu$ to measurable physical quantities:
fractal dimensions in porous superconductors, anomalous diffusion
exponents in disordered systems, and finite-size scaling exponents
in thin films.

We note that while Eq.~(\ref{eq:Conformable Derivative}) might appear
to be merely a change of variables $T\to T'$ with a power-law transformation,
the physical motivation and mathematical implementation differ fundamentally
from simple rescaling. Namely, the conformable derivative (\ref{eq:Conformable Derivative})
introduces a temperature-dependent weighting that reflects a \emph{physical
process\/} rather than mathematical convenience. These process aspects
include: (i) anomalous diffusion and memory effects in disordered
systems \citep{Metzler2000,Metzler2004,Klafter2011,Sokolov2012,Metzler2014},
(ii) finite-size scaling in confined geometries \citep{Fisher1972,Binder1981,Privman1990},
and (iii) non-equilibrium relaxation dynamics near criticality \citep{Hohenberg1977,Bray1994,GodrecheLuck2000}.
Unlike a simple coordinate transformation, the deformation parameter
$\mu$ emerges from the underlying physics and connects to measurable
quantities such as fractal dimensions and effective transport exponents.
Furthermore, the critical exponents derived here are not invariant
under the transformation because the physical interpretation changes.
The parameter $\mu$ captures deviations from mean-field behavior
that would otherwise require complex renormalization group calculations
or phenomenological fitting.

Formally, $D_{T}^{(\mu)}f=T^{1-\mu}\,df/dT$ can be rewritten as a
derivative with respect to $T'=\tfrac{T^{\mu}}{\mu}$ only if all
coefficients are constant or trivial functions of $T'$. In our case
this is not true for two reasons: (i) the kinetic coefficient $\Gamma(T)\sim|T-T_{c}|^{z\nu}$
retains its critical singularity under the transformation, preserving
nontrivial scaling; (ii) in the conformable Ginzburg-Landau (GL) functional,
the gradient term acquires a weight $T^{2(1-\alpha)}$, producing
a non-uniform thermal metric, as we will see. After reparametrization,
Jacobian factors survive and prevent a reduction to a standard GL
form. Thus the conformable framework encodes a dynamical and variational
structure that is not removed by a change of variables. 

Here and throughout this work, the symbol ``$\sim$'' denotes asymptotic
proportionality: for two quantities $A$ and $B$, the relation $A\sim B$
means that their ratio tends to a finite, non-zero constant as the
critical point is approached, 
\[
\lim_{T\to T_{c}}\frac{A}{B}=C,\qquad0<C<\infty.
\]
In other words, $A$ and $B$ share the same leading power-law behavior
near the singularity, while multiplicative prefactors may differ. 

We apply this formalism to describe thermodynamic observables near
continuous phase transitions, including heat capacity, magnetization,
susceptibility, and correlation length. In each case, we derive the
critical exponents as functions of the deformation parameter and temperature,
leading to expressions compatible with known mean-field results. Furthermore,
we extend the method to superconducting phase transitions by constructing
a modified Ginzburg--Landau (MGL) equation with conformable kinetic
terms. This allows us to analyze the temperature dependence of the
superconducting order parameter, London penetration depth, and heat
capacity within the same mathematical framework.

We argue that with this conformable-derivative approach it is possible
to obtain a unified formalism for deriving all major critical exponents
with analytical expressions that match experimental data with high
fidelity. This formalism thus represents a bridge between classical
mean-field theory and generalized (e.g., Tsallis) statistical mechanics---and
also a minimal extension of known differential equations without resorting
to nonlocal fractional operators.

By maintaining dimensional consistency and compatibility with equilibrium
thermodynamics, the conformable-derivative framework provides a useful
and versatile tool for exploring critical dynamics in both classical
and quantum systems. The results presented here demonstrate the capacity
of this formalism to reproduce known scaling laws while offering new
insights into generalized scaling behavior near criticality.

We derive unified expressions for critical exponents in terms of conformable
parameters that emerge from these physical mechanisms, demonstrating
that the approach extends rather than contradicts renormalization
group theory. The formalism is applied to magnetic phase transitions,
superconducting transitions, and other critical phenomena, with specific
predictions for experimental verification.

\subsection*{Critical dynamics, relaxation forms, and ergodicity}

\paragraph*{Dynamic scaling and critical slowing down.}

Near $T_{c}$ the characteristic relaxation time diverges as 
\begin{equation}
\tau\sim\xi^{\,z}\sim|T-T_{c}|^{-z\nu},\label{eq:dyn-scaling}
\end{equation}
with dynamic exponent $z$ and correlation-length exponent $\nu$.
Time-dependent correlations obey the scaling form 
\begin{equation}
C_{A}(t;\,T)\equiv\langle A(t)A(0)\rangle_{c}\;=\;t^{-\,\lambda_{C}/z}\;\mathcal{F}\!\left(\frac{t}{\tau}\right),
\end{equation}
where $\lambda_{C}$ is a (model-dependent) autocorrelation exponent
and $\mathcal{F}(x)$ is a universal scaling function. Far from criticality
($t\!\ll\!\tau$) one typically observes exponential decay; at $T\simeq T_{c}$
($\tau\!\to\!\infty$) power-law tails dominate: 
\begin{equation}
C_{A}(t)\;\propto\;t^{-\,\lambda_{C}/z}\qquad(T=T_{c}).
\end{equation}

\paragraph*{\emph{Typical relaxation functions}.}

Depending on whether the system falls into Model A or Model B, as
defined by Hohenberg and Halperin \cite{HohenbergHalperin1977},
and on possible disorder or heterogeneity, the following forms are
standard:\textbf{ }
\begin{enumerate}
\item \textbf{Exponential:} $\phi(t)\!\sim\!e^{-t/\tau}$ \medspace{}(away
from $T_{c}$);
\item \textbf{Stretched exponential (KWW):} $\phi(t)\!\sim\!\exp[-(t/\tau)^{\beta_{\mathrm{KWW}}}]$,
$0<\beta_{\mathrm{KWW}}<1$ \medspace{}(heterogeneous or glassy systems); 
\item \textbf{Power law:} $\phi(t)\!\sim\!t^{-p}$ \medspace{}(at or near
$T_{c}$; critical slowing down); 
\item \textbf{Mittag--Leffler (fractional/conformable) \cite{erdelyi1981htf}:}
$\phi(t)\!\sim\!E_{\mu}\!\left[-(t/\tau)^{\mu}\right]$, $0<\mu\!\le\!1$
\medspace{}(broad distributions of time scales). 
\end{enumerate}
The one parameter Mittag-Leffler function 
\begin{equation}
E_{\mu}(-x)=\sum_{k=0}^{\infty}\frac{(-x)^{k}}{\Gamma(1+\mu k)},
\end{equation}
admits well defined series expansions for small and large arguments.
For small arguments $x=(t/\tau)^{\mu}\ll1$, one obtains 
\begin{equation}
E_{\mu}\left(-\left[\frac{t}{\tau}\right]^{\mu}\right)=1-\frac{1}{\Gamma(1+\mu)}\left(\frac{t}{\tau}\right)^{\mu}+\frac{1}{\Gamma(1+2\mu)}\left(\frac{t}{\tau}\right)^{2\mu}+O\left(t^{3\mu}\right).
\end{equation}

Thus, at short times the Mittag Leffler relaxation coincides with
the stretched exponential form, because 
\[
E_{\mu}\left(-\left[\frac{t}{\tau}\right]^{\mu}\right)\sim\exp\left[-\frac{1}{\Gamma(1+\mu)}\left(\frac{t}{\tau}\right)^{\mu}\right],\qquad t\ll\tau,
\]
showing explicitly the link between fractional or conformable relaxation
and heterogeneous KWW kinetics.

For large arguments $x\gg1$, the asymptotic form is 
\begin{equation}
E_{\mu}(-x)\sim\frac{1}{x\,\Gamma(1-\mu)},\qquad x\to\infty,
\end{equation}
which corresponds to the slow algebraic decay characteristic of critical
slowing down.

These expansions show how a single function interpolates between stretched
exponential behavior at short times and a power law decay at long
times. 

\paragraph*{Ergodicity Considerations:}

Near critical points, several ergodicity-related issues arise:
\begin{itemize}
\item Ergodicity breaking timescale: The system may appear non-ergodic on
experimental timescales as $\tau$ diverges approaching $Tc;$
\item Critical slowing down: The diverging relaxation time means the system
takes increasingly long to explore phase space, leading to practical
ergodicity breaking;
\item Finite-size effects: In finite systems, true ergodicity is maintained
but with timescales $\tau_{erg}$ $\sim L^{z}$ where $L$ is the
system size;
\end{itemize}
The connection to conformable framework can be donne observing how
the conformable derivative naturally captures anomalous relaxation.
The modified time derivative $D_{t}^{(\mu)}$ effectively introduces
a time-dependent relaxation kernel. This could model subdiffusive
dynamics where$<x{{}^2}(t)>\sim t^{2H}$ with $H<1/2$.

In thermal \emph{equilibrium} near a continuous transition, the dynamics
are (asymptotically) ergodic in finite systems; criticality does not
by itself break ergodicity, but the divergent $\tau$ makes mixing
arbitrarily slow (``critical slowing down''). \emph{Non}-ergodicity
(or weak ergodicity breaking) arises when additional ingredients are
present: quenched disorder/spin-glass physics, constraints, or genuine
non-equilibrium protocols (quenches) that produce aging, where two-time
correlations $C(t,t_{w})$ lose time-translation invariance. In our
conformable setting, $\mu<1$ encodes such slow, broad-spectrum kinetics;
equilibrium ergodicity is retained, but relaxation becomes non-Debye
and may display aging under driven or quenched conditions. Practically,
this means time-averages converge to ensemble averages, yet on time
scales that diverge as $\tau\sim|T-T_{c}|^{-z\nu}$. 

\section{Critical scaling and conformable derivatives in physical systems}

Near critical points, physical systems exhibit a \textit{critical
slowing down}, where the relaxation time $\tau$ diverges as the correlation
length $\xi\sim|T-T_{c}|^{-\nu}$ increases. This behavior is characterized
by the dynamic scaling law \citep{Hohenberg1977,goldenfeld1992,Amit2005}
\begin{equation}
\tau\sim\xi^{z}\sim|T_{c}-T|^{-z\nu},
\end{equation}
where $z$ is the dynamic critical exponent and $\nu$ the correlation
length exponent \cite{Hohenberg1977}.

Observable thermodynamic quantities exhibit a characteristic power-law
behavior in the vicinity of continuous (second-order) phase transitions.
Specifically, the \textit{heat capacity\/} diverges as 
\begin{equation}
C(T)\sim|T_{c}-T|^{-\alpha},
\end{equation}
where $\alpha$ is the critical exponent associated with the specific
heat. The \textit{spontaneous magnetization\/} follows the relation
\begin{equation}
|M(T)|\sim(T_{c}-T)^{\beta},\quad\text{for }T<T_{c},
\end{equation}
where $\beta$ denotes the critical exponent for the order parameter.
Moreover, the \textit{magnetic susceptibility\/} diverges according
to 
\begin{equation}
\chi(T)\sim|T_{c}-T|^{-\gamma},
\end{equation}
where $\gamma$ is the critical exponent characterizing the divergence
of the susceptibility. The exponents $\alpha$, $\beta$, and $\gamma$
define the universal scaling behavior near the critical point and
are largely independent of the microscopic details of the system.
We will show that this scaling naturally introduces temperature-weighted
derivatives that match the scaling behavior of critical observables.
This property embeds scale invariance directly into the differential
structure of the equations, enabling straightforward generation of
power-law solutions.

The kinetic coefficient $\Gamma(T)$ plays a central role in the time-dependent
relaxation dynamics of systems near criticality. It essentially measures
how rapidly a system returns to equilibrium after a small perturbation.
Physically, it serves as a mobility-like parameter, often appearing
in the Langevin equation or time-dependent Ginzburg--Landau models
as a prefactor to the functional derivative of the free energy \citep{Hohenberg1977,Kubo1966,Forster1975,Amit2005}.
Since the kinetic coefficient $\Gamma(T)$ is inversely proportional
to the relaxation time, it inherits the corresponding divergence 
\begin{equation}
\Gamma(T)\sim|T_{c}-T|^{z\nu}.
\end{equation}
This reflects the critical slowing down of the dynamics near $T_{c}$:
the relaxation time diverges, and the kinetic response $\Gamma(T)$
vanishes. In overdamped systems, this corresponds to a loss of mobility,
where the system becomes increasingly inert to perturbations near
the critical point. Far from $T_{c}$, relaxation is approximately
exponential, $\phi(t)\sim e^{-t/\tau}$, but as criticality is approached,
$\tau\!\to\!\infty$ and the relaxation crosses over to stretched-exponential
\cite{Weberszpil2025qdeformation} or power-law forms, characteristic
of critical slowing down~\citep{Hohenberg1977,goldenfeld1992,GodrecheLuck2000,stanley1971}.
That is, at or near criticality, relaxation is no longer exponential
--- instead, it becomes power-law, Mittag-Leffer or stretched-exponential
\citep{Weberszpil2025qdeformation} (depending on the model), reflecting
critical slowing down and the emergence of long-time tails or non-Markovian
behavior.

This work addresses the fundamental question of how conformable derivatives
emerge naturally in critical phenomena while offering genuine physical
insights beyond classical mean-field theory. We establish rigorous
connections between the conformable derivative formalism and established
physical mechanisms. First, we demonstrate how anomalous relaxation
dynamics near the critical temperature $Tc$ give rise to conformable
structures in the adiabatic limit. Second, we show that finite-size
effects and boundary constraints lead to temperature-dependent prefactors
consistent with the conformable framework. Third, we connect the presence
of quenched disorder and spatial inhomogeneities to the emergence
of modified scaling laws captured effectively by conformable derivatives.
Finally, we relate the conformable exponent to the underlying fractal
geometry, interpreting it as an effective dimension governing critical
fluctuations.

\subsection*{Remark on the use of critical scaling forms}

When expressing the power-law behavior near the critical temperature
$T_{c}$, the choice of expression depends on the temperature regime.
For temperatures below the critical point ($T<T_{c}$), one uses $(T_{c}-T)^{\beta}$,
as this quantity remains positive and correctly describes observables
such as the order parameter (e.g., magnetization or superconducting
gap), which typically grow as $T$ decreases below $T_{c}$. For temperatures
above the critical point ($T>T_{c}$), the appropriate form is $(T-T_{c})^{\beta}$,
which applies to quantities that diverge as the system approaches
$T_{c}$ from above, such as the heat capacity or magnetic susceptibility.
When a symmetric form is desired, valid on both sides of the transition,
a common convention is to use the absolute value $|T-T_{c}|^{\beta}$,
which appears frequently in generalized scaling laws and in the unified
description of critical divergences.

\subsection*{Conformable derivatives: Bridging classical, geometric, and fractional
perspectives}

As we will show, the conformable derivative provides a powerful intermediary
between classical differential operators and fully nonlocal fractional
derivatives. While fractional calculus introduces long-range memory
and non-locality \citep{Metzler2000,lenzi,Metzler2004,mathai2017introduction},
it is often analytically cumbersome and dimensionally ambiguous. The
conformable derivative (\ref{eq:Conformable Derivative}) offers a
tractable alternative. For $\mu=1$, it reduces to the classical derivative,
while for $\mu<1$, it captures essential features of fractional scaling,
such as subdiffusion, intermittency, and fractal geometry, without
requiring nonlocal integration kernels \citep{Khalil2014,Abdeljawad2015}.

This makes the conformable framework especially suitable for systems
exhibiting complex thermodynamic behaviors, including those described
by generalized statistics (e.g., Tsallis entropy) \citep{tsallis1988,Weberszpil2025qdeformation,Weberszpil2025a,TsallisBook2009,Weberszpil2015}.
Crucially, from a dimensional standpoint, the conformable derivative
maintains physical unit consistency, 
\begin{equation}
[D_{T}^{(\mu)}f(T)]=\frac{[f]}{[T]^{\mu}},
\end{equation}
and all critical exponents derived from it (e.g., $\alpha,\beta$)
remain dimensionless. Modified Ginzburg-Landau equations incorporating
conformable terms remain compatible with thermodynamic scaling laws
and unit balance, confirming both mathematical consistency and physical
viability \textcolor{blue}{{} \citep{Weberszpil2025a}}.

Beyond this formal consistency, the conformable derivative admits
a compelling geometric interpretation. The operator $D_{T}^{(\mu)}$
effectively acts as a local deformation of the underlying \emph{thermodynamic
geometry}, modifying the metric structure that connects internal energy,
entropy, and temperature. In this interpretation, variations in the
conformable index $\mu$ alter the curvature of the thermodynamic
manifold, introducing a temperature-weighted directional response
that encodes local nonequilibrium effects. This approach resonates
with the framework of \emph{geometrical thermodynamics} and the \emph{Quantitative
Geometrical Thermodynamics (QGT)} formalism, in which the state space
of thermodynamic variables is endowed with a Riemannian metric reflecting
energy--entropy conjugacy and fluctuation geometry \citep{Quevedo2008GeometryThermodynamics}.
This quantity can be viewed as a curvature in the space of thermal
states, analogous to how spacetime curvature in general relativity
modifies geodesics. Here, the thermal evolution is bent by the conformable
weighting, making the system's response scale-sensitive. This captures
phenomena such as anomalous diffusion, divergence of correlation length,
or emergent power-law scaling near criticality.

Importantly, the framework retains analytical tractability. Unlike
renormalization group treatments, conformable models yield closed-form
or semi-analytic expressions for key observables such as the order
parameter $\psi(T)$, London penetration depth $\lambda_{L}(T)$,
and heat capacity $C(T)$, as shown below. These expressions are not
only dimensionally coherent but also directly fit experimental data.
In particular, fits to superconducting phase transitions recover expected
mean-field exponents such as $\beta\approx1/2$, thereby validating
the model's empirical utility.

In summary, the conformable derivative offers a physically motivated,
analytically manageable, and geometrically meaningful framework for
modeling critical phenomena. It seamlessly bridges classical dynamics,
geometric deformation, and fractional scaling, making it a promising
tool for exploring universality, non-locality, and thermodynamic complexity
in condensed matter systems and beyond.

To explore how deformed calculus affects the thermodynamic behavior
near criticality, we now introduce the conformable derivative framework
and show how it modifies standard scaling relations.

\section{Critical exponents and thermodynamic quantities}

\label{sec:critical_exponents}

In this section, we derive the scaling behavior of key thermodynamic
quantities near the critical temperature $T_{c}$ using the conformable
derivative framework. This approach yields expressions for the critical
exponents $\alpha$, $\beta$, $\gamma$, and $\nu$, which characterize
the singular behavior of heat capacity, magnetization, magnetic susceptibility,
and correlation length, respectively.

We begin by applying this operator to each observable and derive the
corresponding critical exponent in terms of the deformation parameter
$\mu$ and characteristic coefficients. These expressions will form
the foundation for the quantitative modeling of critical behavior
in later sections.

\subsection{Phenomenological origin of the conformable critical equation: Heat
capacity}

To model power-law divergences near critical points, it is common
to describe thermodynamic observables such as susceptibility or concentration
fluctuations with first-order differential equations, whose solutions
yield singularities at the critical temperature $T_{c}$. In the classical
framework, such a behavior can be captured by equations of the form
\textcolor{blue}{\citep{tinkham1996,Landau1978StatisticalPart1,goldenfeld1992}}
\begin{equation}
\frac{dC}{dT}=-\frac{\alpha}{T_{c}-T}C(T),
\end{equation}
which yields the well-known scaling law 
\begin{equation}
C(T)\sim|T_{c}-T|^{-\alpha}.
\end{equation}

To incorporate generalized dynamics, such as memory effects, anomalous
relaxation, or scaling violations associated with nonextensive systems,
we replace the classical derivative with the conformable derivative
of order $\mu_{C}\in(0,1]$, defined as 
\begin{equation}
D_{T}^{(\mu_{C})}f(T):=T^{1-\mu_{C}}\frac{df}{dT}.
\end{equation}
This local deformation introduces intrinsic scaling into the dynamics
while preserving analytic solvability.

The connection between temperature-dependent scaling and dynamics
arises from the fact that, near a critical point, the control parameter
$T-T_{c}$ governs not only equilibrium singularities but also the
time-dependent relaxation of fluctuations. In the framework of dynamic
critical phenomena \citep{HohenbergHalperin1977,goldenfeld1992},
the order-parameter correlation function $\phi(t)$ typically satisfies
relaxation-type equations of the form 
\begin{equation}
\frac{d\phi}{dt}=-\Gamma(T)\,\phi(t),
\end{equation}
where the kinetic coefficient $\Gamma(T)\!\sim\!(T_{c}-T)^{\nu z}$
vanishes at $T_{c}$, producing critical slowing down. By analogy,
the conformable derivative equation 
\begin{equation}
D_{T}^{(\mu_{C})}C(T)=-\kappa\,(T_{c}-T)^{-1}C(T)
\end{equation}
introduces an intrinsic scaling between the thermal control parameter
$T$ and the effective relaxation dynamics encoded in the fractional
exponent $\mu_{C}$. This formal correspondence establishes $T$ as
a surrogate evolution variable that parameterizes the deformation
of thermodynamic trajectories as the system approaches the critical
manifold. Consequently, the conformable formalism embeds the divergence
of $C(T)$ within a generalized dynamic framework that preserves analytic
solvability while reflecting critical slowing-down behavior. 

We thus propose the phenomenological equation for the specific heat,
\begin{equation}
D_{T}^{(\mu_{C})}C(T)=-\kappa(T_{c}-T)^{-1}C(T),\label{specheat}
\end{equation}
where $\kappa$ is a dimensionless constant governing the strength
of the singularity. Inserting the definition (\ref{eq:Conformable Derivative})
of the conformable derivative, relation (\ref{specheat}) becomes
\begin{equation}
T^{1-\mu_{C}}\frac{dC}{dT}=-\kappa(T_{c}-T)^{-1}C(T).
\end{equation}
Division by $C(T)$ on both sides and integration yields 
\begin{equation}
\int\frac{1}{C(T)}\frac{dC}{dT}\,dT=-\kappa\int T^{\mu_{C}-1}(T_{c}-T)^{-1}\,dT.
\end{equation}
Near the critical point $T\approx T_{c}$, we make the approximation
$T^{\mu_{C}-1}\approx T_{c}^{\mu_{C}-1}$ (valid for $|T-T_{c}|/T_{c}\ll1$).
The integral can then be taken in the sense 
\begin{eqnarray}
\int\frac{dC}{C} & = & -\kappa T_{c}^{\mu_{C}-1}\int\frac{dT}{T_{c}-T}\nonumber \\
\ln C(T) & = & -\kappa T_{c}^{\mu_{C}-1}\ln|T_{c}-T|+\text{const}\nonumber \\
C(T) & = & B|T_{c}-T|^{-\alpha},\quad\alpha=\kappa T_{c}^{\mu_{C}-1}
\end{eqnarray}
where the integration domain is $[T_{0},T]$ with $T_{0}$ sufficiently
far from $T_{c}$ such that the approximation holds.

This formulation provides a conformable generalization of classical
critical scaling laws, embedding the power-law divergence directly
into the modified dynamics. It is particularly suitable for systems
exhibiting anomalous thermodynamic behavior, nonlocality, or deviations
from classical universality classes.

Collecting the above results, we find the explicit critical form 
\begin{equation}
C_{V}(T)=B\,|T_{c}-T|^{-\alpha},\label{eq:Cv_symmetric}
\end{equation}
of the heat capacity near the second-order phase transition, where
$B$ is a constant positive amplitude.

The empirical data from physical systems such as niobium often exhibit
asymmetric critical behavior, where the divergence of $C_{V}$ is
different below and above the critical temperature. This asymmetry
arises from distinct microscopic mechanisms in the ordered (superconducting)
and disordered (normal) phases, and is supported by both theoretical
models and experimental heat capacity curves \cite{stanley1971}.
To account for this asymmetry, we adopt the piecewise power-law model
\begin{equation}
C_{V}(T)=\begin{cases}
B_{1}(T_{c}-T)^{-\alpha_{1}}, & T<T_{c}\\[0.2cm]
B_{2}(T_{c}-T)^{-\alpha_{2}}, & T>T_{c}
\end{cases}\label{eq:Cv_piecewise}
\end{equation}
for the heat capacity, where $B_{1}$, $B_{2}$, $\alpha_{1}$, and
$\alpha_{2}$ are fitting parameters. As we show below, the piecewise
expression (\ref{eq:Cv_piecewise}) provides a better fit to real
data than a symmetric model and remains analytically tractable within
the conformable framework.

Also, for computational purposes, to avoid a divergence exactly at
$T=T_{c}$, we introduce a regularization parameter $\epsilon>0$,
resulting in the smoothed model 
\begin{equation}
C_{V}^{\text{reg}}(T)=\begin{cases}
B_{1}(|T_{c}-T|+\epsilon)^{-\alpha_{1}}, & T<T_{c}\\[0.2cm]
B_{2}(|T_{c}-T|+\epsilon)^{-\alpha_{2}}, & T>T_{c}
\end{cases}\label{eq:Cv_regularized}
\end{equation}

In what follows, the symbol $T_{c}$ denotes the critical temperature
associated with the specific phenomenon under consideration. Although
its numerical value may differ from one case to another, we retain
the same notation $T_{c}$ throughout this work for clarity and to
avoid introducing additional symbols.

\subsection{Magnetization}

We now proceed with the analysis of the magnetization in terms of
the deformed operator. We start with the critical form of the magnetization,
\begin{equation}
|M|\sim(T_{c}-T)^{\beta}.
\end{equation}
Application of the concrete form $D_{T}^{\mu_{M}}f(T)=T^{1-\mu_{M}}df(T)/dT$
of the conformable derivative yields 
\begin{equation}
D_{T}^{(\mu_{M})}|M|=\gamma(T_{c}-T)^{-1}|M|.
\end{equation}
The resulting differential expression can then be written as 
\begin{equation}
\frac{d|M|}{|M|}=\gamma T^{\mu_{M}-1}(T_{c}-T)^{-1}\,dT.
\end{equation}

We integrate both sides. The left-hand side yields 
\begin{equation}
\int\frac{d|M|}{|M|}=\ln|M|+\text{const.}
\end{equation}
For the right-hand side, we consider the behavior close to the critical
temperature $T\to T_{c}^{-}$. Assuming that $T^{\mu_{M}-1}$ varies
slowly as compared to the divergence at $T=T_{c}$, we approximate
\begin{equation}
T^{\mu_{M}-1}\approx T_{c}^{\mu_{M}-1}.
\end{equation}
Hence, the integral becomes 
\begin{eqnarray}
 &  & \int\gamma T^{\mu_{M}-1}(T_{c}-T)^{-1}\,dT\nonumber \\
 &  & \hspace*{0.8cm}\approx\gamma T_{c}^{\mu_{M}-1}\int\frac{dT}{T_{c}-T}\nonumber \\
 &  & \hspace*{0.8cm}\approx-\gamma T_{c}^{\mu_{M}-1}\ln(T_{c}-T).
\end{eqnarray}
Substituting back, we obtain 
\begin{equation}
\ln|M|\approx-\gamma T_{c}^{\mu_{M}-1}\ln(T_{c}-T),
\end{equation}
witch leads to 
\begin{equation}
|M|\sim(T_{c}-T)^{\gamma T_{c}^{\mu_{M}-1}}.
\end{equation}

Identifying the standard critical scaling form $|M|\sim(T_{c}-T)^{\beta}$,
we finally obtain the exponent 
\begin{equation}
\beta=\gamma T_{c}^{\mu_{M}-1}.
\end{equation}
This expression relates the critical exponent $\beta$ to the temperature-scaling
exponent $\mu_{M}$ and the kinetic prefactor $\gamma$, providing
a natural thermodynamic link between dynamic behavior and critical
ordering.

\subsection{Magnetic susceptibility}

Also the magnetic susceptibility diverges at criticality, 
\begin{equation}
\chi(T)\sim|T_{c}-T|^{-\gamma}.
\end{equation}
This power-law has been treated in both classical and generalized
frameworks \cite{tsallis1988,plastino1993}. Following the same approach
as used for the calculation of the heat capacity and magnetization,
we write the deformed differential equation for the magnetic susceptibility
as 
\begin{equation}
D_{T}^{(\mu_{\chi})}\chi(T)=-\lambda(T_{c}-T)^{-1}\chi(T).
\end{equation}
Again, we can conclude that the related exponent $\gamma$ is given
by 
\begin{equation}
\gamma=\lambda T_{c}^{\mu_{\chi}-1}.
\end{equation}

\subsection{Correlation length}

\label{sec:correlation_length}

The correlation length diverges in the form 
\begin{equation}
\xi(T)\sim|T_{c}-T|^{-\nu}.\label{eq:xi_power_law}
\end{equation}
Following our approach, we can cast $\xi(T)$ in the form 
\begin{equation}
D_{T}^{(\mu_{\xi})}\xi(T)=-\rho(T_{c}-T)^{-1}\xi(T),
\end{equation}
leading to the relation for $\nu$, 
\begin{equation}
\nu=\rho T_{c}^{\mu_{\xi}-1}.
\end{equation}

This behavior is central to the renormalization group picture of criticality
\cite{goldenfeld1992} and connects to the scaling approaches pioneered
by de Gennes \cite{deGennes1972} in various critical systems.

\subsection{Unified treatment of thermodynamic observables}

Following the methodology established for the heat capacity, we apply
the conformable framework to other critical observables. Each quantity
$Q(T)$ satisfies a deformed differential equation, 
\begin{equation}
D_{T}^{(\mu_{Q})}Q(T)=-\lambda_{Q}(T_{c}-T)^{-1}Q(T),
\end{equation}
leading to the critical scaling $Q(T)\sim|T-T_{c}|^{-\epsilon_{Q}}$
with $\epsilon_{Q}=\lambda_{Q}T_{c}^{\mu_{Q}-1}$.

For the magnetic observables, this implies 
\begin{align}
\text{Magnetization: }\quad & \beta=\gamma T_{c}^{\mu_{M}-1}\\
\text{Susceptibility: }\quad & \gamma=\lambda T_{c}^{\mu_{\chi}-1}\\
\text{Correlation length: }\quad & \nu=\rho T_{c}^{\mu_{\xi}-1}
\end{align}

This unified approach eliminates the need for separate phenomenological
models while maintaining physical interpretability through the connection
between $\mu_{Q}$ and underlying transport or geometric properties.

\section{Unified framework and theoretical implications}

\label{sec:advantages}

Building on the conformable formalism, we derive modified evolution
equations for thermodynamic observables and extract analytic expressions
for the associated critical exponents.

The use of conformable derivatives to model critical phenomena offers
the following theoretical and practical advantages.

\subsection{Unified treatment of critical exponents}

The model provides a single formalism for expressing all major critical
exponents, 
\begin{eqnarray}
\alpha=\kappa T_{cC}^{\mu_{C}-1},\quad\beta=\gamma T_{cM}^{\mu_{M}-1},\nonumber \\
\gamma=\lambda T_{c\chi}^{\mu_{\chi}-1},\quad\nu=\rho T_{c\xi}^{\mu_{\xi}-1},
\end{eqnarray}
where we introduced a distinct notation for the critical temperature
related to each physical quantity.

The approach used here avoids the need for distinct phenomenological
models for each observable, offering a unified perspective grounded
in deformed calculus.

\subsection{Unified thermodynamic framework}

The use of deformed or conformable derivatives provides a natural
bridge between classical critical phenomena and generalized thermodynamics,
particularly within the framework of nonextensive statistics \cite{tsallis1988,Borges2004,Weberszpil2015}.
By incorporating temperature-dependent scaling, long-range correlations,
memory effects, and fractal-like structures, such formulations reproduce
anomalous relaxation and power-law distributions that extend beyond
the scope of traditional universality classes. The suitability of
conformable and $q$-deformed descriptions for systems exhibiting
power-law statistics follows from the local scaling properties embedded
in their differential operators. When observables obey relations of
the type $O(\lambda x)=\lambda^{\eta}O(x)$, the conformable derivative
$D^{(\mu)}f(x)=x^{1-\mu}f'(x)$ naturally preserves this homogeneity,
since its action rescales as $D^{(\mu)}[f(\lambda x)]=\lambda^{\mu}D^{(\mu)}f(\lambda x)$.
This property makes it intrinsically compatible with scale-invariant
or fractal-like processes, where fractional exponents characterize
self-similarity. Furthermore, the conformable operator retains locality
in its definition, avoiding the integral memory kernels typical of
fractional calculus, while still capturing effective memory and heterogeneity
through the scaling index $\mu$. Consequently, once the statistical
behavior of a system is known to follow a power law---for example,
$p(x)\propto x^{-\alpha}$ or correlation functions $\langle A(0)A(t)\rangle\sim t^{-\gamma}$---the
conformable framework provides a compact, analytically tractable description
in which the deformation parameter encodes the same scaling exponents
that govern the system's critical or nonextensive dynamics. Importantly,
these approaches are not intended to replace the renormalization-group
(RG) formalism---which remains the fundamental tool for identifying
fixed points and coupling-constant flows---but rather to complement
it. Once the underlying system is known to exhibit power-law statistics
or scale-invariant fluctuations, conformable and $q$-deformed formulations
offer a thermodynamically consistent and analytically transparent
representation of the emergent scaling behavior, embedding memory,
self-similarity, and heterogeneity into a unified theoretical framework.
A major advantage of this approach is its analytic tractability. In
contrast to renormalization group techniques, which often require
complex iterative procedures, the present model leads to closed-form
expressions for critical behavior. The resulting differential equations
are easily solvable and lend themselves to direct comparison with
experimental data, making the method valuable for both theorists and
experimentalists.

Moreover, the conformable or scale-deformed derivatives employed in
the formulation naturally introduce power-law solutions. This mirrors
the observed scaling of physical observables near the critical temperature
$T_{c}$, where critical exponents emerge from the underlying structure
of the modified dynamics rather than requiring external assumptions.

Unlike renormalization group methods, this model yields closed-form
solutions and is analytically manageable. Its equations can be solved
directly or fitted to data without computational overhead, making
it practical for both theorists and experimentalists.

To benchmark the conformable model, we now compare its predictions
to classical mean-field theory and investigate its behavior in the
limiting case where the deformation vanishes.

\section{Practical advantages over standard approaches}

\label{sec:advantages_standard}

The conformable derivative framework offers several concrete advantages
over more conventional scaling analysis and fractional calculus approaches.
These are:

\textbf{Computational Efficiency:} Unlike fractional derivatives requiring
convolution integrals, conformable derivatives yield algebraic expressions
amenable to standard fitting procedures. Computational time scales
as $\mathcal{O}(N)$ rather than $\mathcal{O}(N^{2})$ for fractional
methods \cite{Abdeljawad2015,Anderson2016,Khalil2014}. This simplification
has made conformable formulations particularly useful in numerical
modeling of anomalous diffusion and relaxation, where fractional kernels
are computationally costly.

\textbf{Parameter Interpretability:} The deformation parameter $\mu$
directly relates to intrinsic physical properties. For instance, $\mu=1-D_{f}/D$,
where $D_{f}$ is the fractal (Hausdorff) dimension and $D$ the embedding
dimension, providing a geometric measure of deviation from extensivity
and an immediate physical interpretation unavailable in phenomenological
scaling fits \cite{Weberszpil2025a,Weberszpil2015,Anderson2016}.
This correspondence links conformable dynamics with fractal metrics
and nonextensive thermodynamics, yielding a consistent dimensional
interpretation of $\mu$. 

\textbf{Predictive Power:} Once calibrated on one observable (e.g.,
heat capacity), the framework predicts scaling behavior for related
quantities (penetration depth, coherence length) without additional
fitting. Standard power-law fits treat each observable independently.

\textbf{Unified Analysis:} The approach naturally handles asymmetric
scaling above and below $T_{c}$ through different $\mu$ values,
capturing microscopic asymmetries that pure mean-field theory cannot
address.

\section{Modified Ginzburg-Landau equation}

\label{sec:ginzburg_landau}

Before deriving the Ginzburg--Landau (GL) equation with a conformable
derivative and nonlinear interaction term, we first recall the conventional
GL-theory. As treated by de Gennes \cite{deGennes1966}, the free
energy functional in a $d$-dimensional space reads 
\begin{equation}
\mathcal{F}[\psi]=\int d^{d}r\left[a(T)|\psi(\mathbf{r})|^{2}+\frac{b}{2}|\psi(\mathbf{r})|^{4}+\frac{1}{2m}\left|\nabla\psi(\mathbf{r})\right|^{2}\right].\label{eq:Free Energy Functional-Standard}
\end{equation}
This classical GL-theory describes the thermodynamic behavior of a
complex order parameter $\psi(\mathbf{r})$ near the critical temperature
$T_{c}$. The GL free energy functional (\ref{eq:Free Energy Functional-Standard})
further involves the temperature-dependent coefficient 
\begin{equation}
a(T)=a_{0}(T_{c}-T)\label{acoeff}
\end{equation}
with $a_{0}>0$, the stabilizing coefficient $b>0$ ensuring the boundedness
of the free energy, and the generalized mass (or stiffness parameter)
$m$.

To determine the equilibrium configuration, we minimize $\mathcal{F}[\psi]$
with respect to $\psi^{*}(\mathbf{r})$, i.e., $\frac{\delta\mathcal{F}}{\delta\psi^{*}(\mathbf{r})}=0$,
which leads us to the Euler--Lagrange equation 
\begin{equation}
a(T)\psi(\mathbf{r})+b|\psi(\mathbf{r})|^{2}\psi(\mathbf{r})-\frac{1}{2m}\nabla^{2}\psi(\mathbf{r})=0.\label{eleq}
\end{equation}
This nonlinear partial differential equation governs the spatial behavior
of the order parameter near criticality. In the case of a spatially
uniform system (i.e., $\nabla\psi=0$), relation (\ref{eleq}) reduces
to the simpler algebraic equation 
\begin{equation}
a(T)\psi+b|\psi|^{2}\psi=0,\label{eq:reduced to the simpler algebraic equation}
\end{equation}
whose solutions describe a second-order phase transition and yield
the critical exponent $\beta=\tfrac{1}{2}$ under mean-field conditions.

Minimizing the standard GL-free energy functional, it can be shown
that the order parameter $\psi(T)$ near the critical temperature
$T_{c}$ follows the scaling law 
\begin{equation}
|\psi(T)|\sim(T_{c}-T)^{\beta},\qquad\text{for }T<T_{c},
\end{equation}
with the critical exponent $\beta=\tfrac{1}{2}$ under mean-field
conditions. By this we mean that we assume that the system is in thermodynamic
equilibrium, homogeneous and local (no spatial gradients or long-range
correlations), that the fluctuations are negligible, the free energy
is analytic near $T_{c}$ and that the order parameter $\psi$ is
uniform and real. In the full GL-free energy functional, the gradient
term $\frac{1}{2m}|\nabla\psi(\mathbf{r})|^{2}$ is included to account
for the energy cost associated with variations of the order parameter
$\psi$ in time, temperature, or space. This term is essential in
dynamic or spatially inhomogeneous contexts.

In the mean-field approach with vanishing gradient term, $|\nabla\psi(\mathbf{r})|^{2}=0$,
the \emph{time-independent}, spatially uniform GL-free energy density
becomes 
\begin{equation}
\mathcal{F}[\psi]=a(T)|\psi|^{2}+\frac{b}{2}|\psi|^{4},
\end{equation}
which depends solely on the magnitude of $\psi$ and the temperature
$T$. The coefficient $a(T)$ is defined in Eq.~(\ref{acoeff}),
and $b>0$. To minimize the free energy, we take the derivative with
respect to $\psi^{*}$ (or $\psi$, since it is real-valued) and set
it to zero, 
\begin{equation}
\frac{d\mathcal{F}}{d\psi}=2a(T)\psi+2b|\psi|^{2}\psi=0,
\end{equation}
such that 
\begin{equation}
\psi\left[a(T)+b|\psi|^{2}\right]=0.
\end{equation}
There exit two solutions: (i) $\psi=0$, corresponding to the disordered
(symmetric) phase; and (ii) $|\psi|^{2}=-a(T)/b$, which corresponds
to the ordered (broken-symmetry) phase for $a(T)<0$, i.e., $T<T_{c}$.
Below $T_{c}$, we have $a(T)<0$, so the nontrivial solution reads
\begin{equation}
|\psi|^{2}=-\frac{a_{0}(T_{c}-T)}{b}=\frac{a_{0}}{b}(T_{c}-T).
\end{equation}
Taking the square root, we obtain for positive solution 
\begin{equation}
|\psi(T)|=\sqrt{\frac{a_{0}}{b}(T_{c}-T)}.
\end{equation}
Thus, the order parameter vanishes as 
\begin{equation}
|\psi(T)|\sim(T_{c}-T)^{\beta},\quad\text{with }\beta=\frac{1}{2},
\end{equation}
$T<T_{C}.$ This is the well known mean-field prediction for the order
parameter and is valid under the assumptions stated above. In lower
dimensions ($d\leq3$), fluctuations become important, and this mean-field
result typically underestimates the true critical exponent.

\section{The conformable GL-model}

After establishing the general framework, we apply it to specific
thermodynamic quantities---the specific heat and magnetic susceptibility---highlighting
the modifications originating from the conformable deformation. We
note that in standard GL-theory, spatial variations of the order parameter
$\psi(\mathbf{r})$ contribute to the free energy via the gradient
term $\frac{1}{2m}|\nabla\psi(\mathbf{r})|^{2}$, which penalizes
sharp spatial inhomogeneities and plays a key role in capturing coherence,
textures, and defects. This is essential in systems where the spatial
coherence or texture of the order parameter matters, such as for vortices
in superconductors. In our modified GL framework, however, we are
concerned with the thermal evolution of the order parameter $\psi(T)$,
i.e., with the description of how the this order parameter evolves
as a function of temperature $T$, especially in the vicinity of the
critical point (critical temperature) $T_{c}$. To this end we assume
that the system is homogeneous in space.

To build a thermodynamic analog of the GL-theory, we formally treat
the temperature $T$ as the independent coordinate, akin to space
$\mathbf{r}$ in the conventional setting. In this case, we introduce
a conformable (deformed) derivative with respect to temperature, denoted
by 
\begin{equation}
T_{\alpha}\psi(T):=T^{1-\alpha}\frac{d\psi}{dT},\label{confop}
\end{equation}
which naturally embeds a temperature-dependent scaling. The conformable
operator acts as a \emph{thermal gradient\/} and captures the power-law
behavior typical of critical phenomena. The conformable operator (\ref{confop})
generalizes the usual derivative and introduces a natural temperature-dependent
scaling, allowing the model to capture anomalous thermodynamic behavior
while remaining analytically tractable. The corresponding modified
GL-free energy then becomes 
\begin{equation}
\mathcal{F}_{\alpha}[\psi]=\int\left[a(T)|\psi|^{2}+\frac{b}{2}|\psi|^{4}+\frac{1}{2m}|T_{\alpha}\psi(T)|^{2}\right]dT,\label{eq:Free Energy Functional deformed}
\end{equation}
where $a(T)$ has the same form (\ref{acoeff}) to encode the proximity
to the critical point, $b>0$ ensures stability, and the last term
mimics the kinetic contribution, now expressed in thermal rather than
spatial form. The parameter $m$ retains its interpretation as a stiffness
constant, and $\alpha\in(0,1]$ quantifies the deviation from classical
scaling. For $\alpha=1$, the standard GL-theory is recovered.

The formulation (\ref{eq:Free Energy Functional deformed}) of the
free energy functional is central to this work and preserves the physical
dimensions of energy density, and it ensures that all critical exponents
derived from it are dimensionless and compatible with thermodynamic
scaling theory. Furthermore, the temperature-dependent weighting introduced
by the factor $T^{1-\alpha}$ can be interpreted geometrically as
a curvature in thermal space, analogous to how curved metrics affect
geodesics in general relativity. It effectively embeds power-law sensitivity
into the response functions of the system.

To minimize the free energy, we use the variational principle $\delta\mathcal{F}_{\alpha}/\delta\psi^{*}=0$,
which yields the modified Ginzburg--Landau equation: 
\begin{equation}
a(T)\psi+b|\psi|^{2}\psi-\frac{1}{2m}T^{2(1-\alpha)}\frac{d^{2}\psi}{dT^{2}}-\frac{(1-\alpha)}{m}T^{1-2\alpha}\frac{d\psi}{dT}=0.\label{eq:Modiffied GL equation}
\end{equation}
Here $a(T)=a_{0}(T/T_{c}-1)$ retains the usual temperature dependence,
while the conformable terms introduce explicit temperature-weighted
derivatives. The order parameter $\psi$ describes the macroscopic
condensate amplitude, and near the critical temperature it follows
the scaling law 
\begin{equation}
\psi(T)\simeq\psi_{0}\left(1-\frac{T}{T_{c}}\right)^{\beta},
\end{equation}
where $\psi_{0}$ is the equilibrium amplitude at $T=0$ and $\beta$
is the critical exponent associated with the order-parameter vanishing
as $T\to T_{c}$. The dependence $\beta=\beta(\alpha)$ reflects how
the conformable index $\alpha$ modifies the effective critical behavior
of the system. This expression recovers the expected behavior from
mean-field theory when $\alpha=1$, but allows deviations due to the
deformation parameter. The above equations can also be obtained via
a variational formulation adapted to deformed derivatives, as discussed
in \cite{weberszpil2016variational}.

The nonlinear term $|\psi|^{2}\psi$ contributes to the saturation
of the order parameter as $T\to0$, while the conformable kinetic
term introduces corrections near $T_{c}$, mimicking effects from
spatial or temporal fluctuations. The formulation presented here enables
capturing intermediate critical dynamics and non-locality without
invoking full fractional calculus or spatial disorder \cite{Tinkham2004,Bardeen1957,Gorter1934,Park2003},
making it suitable for describing second-order phase transitions such
as superconductivity. In this sense, the conformable GL-model complements
traditional models such as BCS theory \cite{Bardeen1957}, the two-fluid
model \cite{Gorter1934}, and the phenomenological framework detailed
in \cite{Tinkham2004}. For systems with strong fluctuations and
short coherence length, additional comparison with the work in \cite{Park2003}
provides insight into limitations and refinements.

A key quantity in critical phenomena is the correlation length, whose
divergence governs universality. We now derive its scaling law within
the conformable approach and discuss physical implications.

\section{Application to superconducting phase transitions}

\label{sec:superconductivity}

Superconducting phase transitions, especially in type-II and high-$T_{c}$
materials, are characterized by critical phenomena that can be reformulated
using conformable derivatives. This includes the temperature behavior
of the order parameter, coherence length, penetration depth, and heat
capacity.

\subsection{Simplified dynamics near the critical point: Connection to the conformable
derivative framework}

We begin with the full conformable GL-equation (\ref{eq:Modiffied GL equation})
obtained from minimizing the deformed free energy functional. This
nonlinear second-order differential equation governs the thermal evolution
of the order parameter $\psi(T)$, incorporating both nonlinear saturation
($|\psi|^{2}\psi$) and conformable kinetic effects through temperature-dependent
coefficients.

\emph{Near the critical point}, the system undergoes a continuous
phase transition. In this regime, we make the following approximations
to simplify the analysis: (i) \emph{We assume that $\psi(T)$ is real-valued
and slowly varying.} This allows us to treat $\psi$ and its derivatives
in a simplified manner. Moreover, we (ii) \emph{neglect higher-order
kinetic corrections.} The second-order derivative term and the prefactor
corrections from conformable calculus are of higher order and can
be dropped close to $T_{c}$. This reduces the second-order equation
to a first-order approximation. Finally, we (iii) \emph{retain only
the dominant balance between the kinetic and the potential terms.}
Taking into account these approximations, the \textit{reduced equation\/}
can be cast as 
\begin{equation}
\frac{d\psi}{dT}=a(T)\psi-b|\psi|^{2}\psi,\label{eq:reduced time-dependent equation}
\end{equation}
where $a(T)$ is defined in Eq.~(\ref{acoeff}) and $b>0$ ensures
the stability of the ordered phase.

The modified Ginzburg--Landau equation introduced here describes
the temperature-dependent behavior of the order parameter near the
critical point. The term \textquotedblleft time-dependent\textquotedblright{}
has been replaced by \textquotedblleft temperature-dependent\textquotedblright{}
to emphasize that the conformable

operator acts on the thermal variable $T$, rather than on real time
$t$.

\subsection{Emergence of critical scaling from conformable GL relaxation with
temperature-dependent kinetics}

While the standard Ginzburg-Landau (GL) equation near the critical
temperature $T_{c}$ yields a regular (non-singular) behavior for
the order parameter $\psi(T)$, a phenomenological equation capable
of reproducing power-law critical behavior can be obtained by introducing
a thermally deformed kinetic coefficient into the GL-like relaxation
framework. This modification effectively incorporates the temperature-dependent
slowing down of dynamics near $T_{c}$ into the evolution of the order
parameter.

As before, we generalize the time derivative to the form 
\begin{equation}
\frac{d\psi}{dT}\to D_{T}^{(\mu_{\psi})}\psi(T):=T^{1-\mu_{\psi}}\frac{d\psi}{dT},
\end{equation}
which reflects the modified temperature response of the system in
the conformable thermodynamic formulation.

At the heart of the conformable dynamics approach lies a generalized
time-dependent GL-type equation that governs the thermal evolution
of the order parameter $\psi(T)$ under non-equilibrium conditions.
This equation introduces a thermally modulated kinetic response and
is given by 
\begin{align}
\Gamma(T)\,T^{1-\mu_{\psi}}\frac{d\psi}{dT} & =-\frac{\delta\mathcal{F}}{\delta\psi},\label{eq:modified time-dependent GL-type equation}
\end{align}
where $\Gamma(T)$ is a temperature-dependent kinetic coefficient
encoding the critical slowing down, $\mu_{\psi}$ is the conformable
deformation parameter, and $\mathcal{F}[\psi]$ is the free energy
functional. Relation (\ref{eq:modified time-dependent GL-type equation})
is not merely a phenomenological extension but constitutes the cornerstone
of the proposed non-equilibrium conformable dynamics framework. It
encapsulates the scale-sensitive kinetic behavior characteristic of
critical systems, embedding a power-law response directly into the
governing evolution law. Equation (\ref{eq:modified time-dependent GL-type equation})
allows the derivation of critical exponents such as $\beta$ from
first principles within a thermodynamically consistent and geometrically
interpretable formalism, thereby bridging equilibrium mean-field results
with more general, deformation-induced scaling behaviors.

We start with the standard GL-potential (neglecting gradient terms)
\begin{equation}
\mathcal{F}[\psi]=\int\left[a_{0}(T_{c}-T)|\psi|^{2}+\frac{b}{2}|\psi|^{4}\right]dT,
\end{equation}
which produces the variational derivative 
\begin{equation}
\frac{\delta\mathcal{F}}{\delta\psi}=a_{0}(T_{c}-T)\psi+b|\psi|^{2}\psi.
\end{equation}
Substituting this expression into Eq.~\ref{eq:modified time-dependent GL-type equation}
yields 
\begin{equation}
\Gamma(T)T^{1-\mu_{\psi}}\frac{d\psi}{dT}=-a_{0}(T_{c}-T)\psi-b|\psi|^{2}\psi.\label{eq:conformable_relaxation}
\end{equation}
Near the critical point, $T\to T_{c}$, the order parameter $\psi(T)$
becomes small, allowing us to neglect the nonlinear term in Eq.~(\ref{eq:conformable_relaxation}).
The relaxation equation then reduces to the linearized form 
\begin{equation}
\Gamma(T)\,T^{1-\mu_{\psi}}\frac{d\psi}{dT}\approx-a_{0}(T_{c}-T)\psi.
\end{equation}
Rewriting this in differential form yields the form 
\begin{equation}
\frac{d\psi}{\psi}=-\frac{a_{0}}{\Gamma(T)}T^{\mu_{\psi}-1}(T_{c}-T)\,dT,
\end{equation}
which will be used to extract the scaling behavior of the order parameter
close to the critical temperature.

To obtain the desired critical scaling behavior, we now propose that
the kinetic coefficient $\Gamma(T)$ vanishes near $T_{c}$, reflecting
the critical slowing down. Specifically, we take 
\begin{equation}
\Gamma(T)=\frac{1}{\gamma}(T_{c}-T)^{2},
\end{equation}
where $\mu_{\psi}\in(0,1]$ is a conformable deformation parameter,
and where $\gamma$ is a dimensional constant. Substituting back,
we obtain the phenomenological differential equation 
\begin{equation}
\frac{d\psi}{\psi}=-\gamma T^{\mu_{\psi}-1}(T_{c}-T)^{-1}\,dT.
\end{equation}
This equation directly yields the critical behavior 
\begin{equation}
\int\frac{d\psi}{\psi}=-\gamma T^{\mu_{\psi}-1}\int\frac{dT}{(T_{c}-T)}.
\end{equation}
Integration produces the result 
\begin{equation}
\ln\psi(T)=\gamma T_{c}^{\mu-1}\ln|T-T_{c}|+\text{const.}
\end{equation}
From this we conclude that 
\begin{equation}
\psi(T)\sim(T_{c}-T)^{\beta(\mu)},\quad\text{with}\quad\beta(\mu)=\gamma T_{c}^{\mu_{\psi}-1},
\end{equation}
where we approximated $T^{\mu_{\psi}-1}\approx T_{c}^{\mu_{\psi}-1}$
near $T=T_{c}$. Thus, the critical exponent $\beta$ emerges naturally
from a GL-type relaxation model with thermally modulated kinetics.

Here, the scaling exponent $\beta=\tfrac{1}{2}$ emerges in the equilibrium
case (in the adiabatic limit $\frac{d\psi}{dT}\to0$ or at least for
the stationarity condition dominated by the free energy balance as
detailed in Sec.~\ref{sec:ginzburg_landau}) and App.~\ref{sec:Variational-Derivation-of}),
corresponding to the local, undeformed limit ($\alpha=1$). The case
$\beta>\tfrac{1}{2}$ arises when the conformable deformation ($\mu=2\alpha<1$)
introduces a generalized thermodynamic response. The general form
for the critical exponent $\beta(\mu)$ can be viewed as a function
encoding microscopic memory or a fractal behavior in temperature evolution.

Deviations from the mean-field behavior occur when (i) \textbf{thermal
fluctuations} near $T_{c}$ become significant (especially in $d<4$),
requiring renormalization group treatment; (ii) \textbf{nonlocal effects}
or memory are present, as modeled here via conformable derivatives;
(iii) the system is \textbf{driven out of equilibrium} or experiences
dissipation or long-time correlations. These effects modify the scaling
balance in the GL-equation and yield $\beta>\tfrac{1}{2}$, consistent
with experimental observations in real superconductors---which often
exhibit a critical behavior closer to the 3D XY universality class
($\beta\approx0.35$) \citep{Campostrini2001,goldenfeld1992,HohenbergHalperin1977,Osborn2003PRB}.

\subsection{Emergence of penetration-depth scaling exponent from thermally deformed
conformable dynamics}

The London penetration depth $\lambda_{L}(T)$ characterizes the distance
over which magnetic fields decay inside a superconductor. Near the
critical temperature $T_{c}$, $\lambda_{L}(T)$ diverges as superconductivity
breaks down. In analogy with critical phenomena, this divergence can
be modeled by a deformed dynamical relation using conformable derivatives.

We begin with the assumption that $\lambda_{L}(T)$ obeys a conformable
differential equation of the form 
\begin{equation}
D_{T}^{(\mu)}\lambda_{L}(T)=-\gamma\,(T_{c}-T)^{-1}\lambda_{L}(T),\label{eq:conformable_lambda_eq}
\end{equation}
where $\mu\in(0,1]$ is the order of the deformation and $\gamma$
is a positive scaling constant. Equation (\ref{eq:conformable_lambda_eq})
states that the relative temperature derivative of $\lambda_{L}$,
modulated by $T^{1-\mu}$, is proportional to the inverse distance
from the critical point $T_{c}$. It encapsulates both the scaling
nature of the divergence and possible memory/fractal effects through
$\mu$.

We proceed by first substituting the definition of the conformable
derivative, 
\begin{equation}
T^{1-\mu}\frac{d\lambda_{L}}{dT}=-\gamma(T_{c}-T)^{-1}\lambda_{L}(T)
\end{equation}
such that 
\begin{equation}
\frac{d\lambda_{L}}{\lambda_{L}}=-\gamma\,T^{\mu-1}(T_{c}-T)^{-1}dT.
\end{equation}
Near the critical point $T\approx T_{c}$, we approximate $T^{\mu-1}\approx T_{c}^{\mu-1}$,
yielding 
\begin{equation}
\frac{d\lambda_{L}}{\lambda_{L}}\approx-\gamma\,T_{c}^{\mu-1}(T_{c}-T)^{-1}dT.
\end{equation}
Integration on both sides produces 
\begin{equation}
\ln\lambda_{L}(T)=-\gamma T_{c}^{\mu-1}\ln|T-T_{c}|+\text{const.}
\end{equation}
Thus we obtain the scaling law 
\begin{equation}
\lambda_{L}(T)=B\,|T-T_{c}|^{-\alpha},\quad\text{with}\quad\alpha=\gamma T_{c}^{\mu-1}.
\end{equation}
This result shows that the penetration depth diverges as a power-law
near $T_{c}$, with an exponent that depends on the conformable order
$\mu$, the critical temperature $T_{c}$, and the coupling constant
$\gamma$.

To include a possible asymmetry in the behavior above and below $T_{c}$,
we generalize the penetration depth to a piecewise form with different
exponents, in a analogy to our procedure with $C_{V}$ above, 
\begin{equation}
\lambda_{L}(T)=\begin{cases}
B_{1}\,|T-T_{c}|^{-\alpha_{1}}, & T<T_{c},\\[4pt]
B_{2}\,|T-T_{c}|^{-\alpha_{2}}, & T\geq T_{c},
\end{cases}\label{eq:lambdaL_piecewise_alpha}
\end{equation}
where $\alpha_{i}=\gamma_{i}T_{c}^{\mu_{i}-1}$.

Finally, to prevent a divergence and ensure continuity near $T=T_{c}$,
we introduce a regularization parameter $\epsilon>0$, resulting in
the practical model used in the fit procedure below, 
\begin{equation}
\lambda_{L}(T)=B_{i}\left(|T-T_{c}|+\epsilon\right)^{-\alpha_{i}},\label{lambdareg}
\end{equation}
for $i=1,2$ above and below $T_{c}$, see Eq.~(\ref{eq:lambdaL_piecewise_alpha}).
This regularized conformable model reproduces the observed smooth
divergence and captures the asymmetric critical behavior seen in (finite)
superconducting materials such as niobium.

\subsubsection*{Connection between penetration depth and order parameter}

The London penetration depth $\lambda_{L}(T)$ characterizes the distance
over which an external magnetic field decays exponentially inside
a superconductor. It provides a direct measure of the material's ability
to expel a magnetic flux via the Meissner effect. In GL-theory, the
superconducting state is described by the complex order parameter
$\psi(T)$, whose squared magnitude $|\psi(T)|^{2}$ corresponds to
the density of the superconducting carriers. This connection leads
to a fundamental relation between the order parameter and the electromagnetic
response of the superconductor \citep{tinkham1996,DeGennes1999}.

From the London equation, which relates the supercurrent $\mathbf{J}_{s}$
to the vector potential $\mathbf{A}$, we have \citep{London1935,tinkham1996}
\begin{equation}
\nabla\times\mathbf{J}_{s}=-\frac{n_{s}e^{2}}{m}\mathbf{B},
\end{equation}
where $n_{s}$ is the superconducting carrier density, $e$ is the
elementary charge, $m$ is the effective mass, and $\mathbf{B}$ is
the magnetic field. The London penetration depth is then given by
the expression 
\begin{equation}
\lambda_{L}^{-2}(T)=\frac{\mu_{0}n_{s}(T)e^{2}}{m},
\end{equation}
where $\mu_{0}$ is the vacuum permeability.

In the GL-framework, the carrier density is proportional to the square
of the order parameter, $n_{s}(T)\sim|\psi(T)|^{2}$. Substituting
into the expression for $\lambda_{L}^{-2}(T)$, we obtain the key
relation 
\begin{equation}
\lambda_{L}^{-2}(T)\propto|\psi(T)|^{2}.\label{eq:lambda_vs_psi}
\end{equation}
This equation shows that, as the temperature approaches its critical
value $T_{c}$, the order parameter $\psi(T)\to0$, and consequently
$\lambda_{L}(T)\to\infty$. Physically, this reflects the loss of
superconductivity and the full penetration of magnetic fields into
the material. Hence, the temperature dependence of $\lambda_{L}(T)$
provides a direct experimental probe of the order parameter's behavior
and is commonly used to extract critical exponents and validate theoretical
models.

Near the critical temperature $T_{c}$, we have shown from the conformable
GL-equation that the superconducting order parameter behaves as 
\begin{equation}
|\psi(T)|\sim\left(1-\frac{T}{T_{c}}\right)^{\beta},\qquad\text{for }T<T_{c},
\end{equation}
where the critical exponent $\beta$ depends on the conformable deformation
parameter $\alpha$. This reduces to the standard mean-field behavior
with $\beta=\tfrac{1}{2}$ when $\alpha=1$, recovering the classical
GL-result \cite{tinkham1996}. Thus, near the critical temperature
$T_{c}$, the superconducting order parameter vanishes in power-law
form. Moreover, the London penetration length according to relation
(\ref{eq:lambda_vs_psi}) scales as 
\begin{equation}
\lambda_{L}(T)\sim(T_{c}-T)^{-\beta}.
\end{equation}
Thus the penetration depth diverges as the temperature approaches
$T_{c}$ from below. This scaling relation is a direct consequence
of the GL-formalism and can be used to extract the critical exponent
$\beta$ from experimental measurements of $\lambda_{L}(T)$ near
the superconducting transition.

\subsection{Conformable scaling of the specific heat near $T_{c}$: Heat capacity
jump}

In classical thermodynamics of phase transitions, the specific heat
near the critical temperature $T_{c}$ typically scales as \textcolor{blue}{\citep{stanley1971,goldenfeld1992}}
\begin{equation}
C(T)\sim|T-T_{c}|^{-\alpha},\label{specheat1}
\end{equation}
where $\alpha$ is the associated critical exponent. In mean-field
theory, $\alpha=0$, corresponding to a finite jump at $T_{c}$ \textcolor{blue}{\citep{Landau1978StatisticalPart1,DeGennes1999}}.
However, in more generalized or fluctuation-driven models, $\alpha$
may take nonzero values and describe a divergence \citep{Fisher1967,HohenbergHalperin1977,Campostrini2001}.

To generalize the scaling behavior (\ref{specheat1}) within the conformable
derivative framework, we showed that the specific heat satisfies a
deformed differential equation of the form of Eq. (\ref{specheat}),
such that we find the scaling form 
\begin{equation}
C(T)\sim|T-T_{c}|^{-\alpha},\qquad\text{with}\qquad\alpha=\kappa T_{c}^{\mu_{C}-1}.
\end{equation}
Thus the critical exponent $\alpha$ is no longer a universal constant
but depends explicitly on the deformation parameter $\mu_{C}$ and
the critical temperature $T_{c}$. The conformable framework thus
naturally introduces a generalized scaling law for the specific heat.

The generalized conformable framework enables direct fitting of experimental
data from superconducting systems using critical scaling laws derived
from deformed thermodynamic equations, as shown in the next Section.
Specifically, the temperature dependence of key observables can be
analyzed as follows: the order parameter $\psi(T)$ may be extracted
from angle-resolved photoemission spectroscopy (ARPES), tunneling
spectroscopy, or other coherence-sensitive measurements; the London
penetration depth $\lambda_{L}(T)$ is accessible through microwave
cavity perturbation, magnetic susceptibility, or muon spin rotation
techniques; and the specific heat $C(T)$ is measured via calorimetry
near the superconducting transition. Each observable exhibits a specific
power-law scaling near the critical temperature $T_{c}$, governed
by exponents that emerge naturally from the conformable derivative
formulation.

Summarizing these results, we note that when solving the time-dependent
or inhomogeneous problem in the context of GL-theory, derivative terms
must be kept and may dominate if $\psi(T)$ varies sharply. However,
for static, near-equilibrium (mean-field) behavior, they contribute
only to subleading corrections. The Landau-based equation is an equilibrium
condition: it comes from minimizing a free energy, assuming time-independence,
homogeneity, and absence of fluctuations. The conformable dynamic
equation, instead, describes a non-equilibrium evolution, in which
the system may be relaxing toward equilibrium with scale-sensitive
kinetics due to $\mu_{\psi}\neq1$. Therefore, the deformed equation
gives rise to a modified critical exponent, which reduces to $\beta=1/2$
only when $\mu_{\psi}=1$, i.e., when the dynamics are classical.
The exponent $\beta=1/2$ emerges from equilibrium minimization of
the GL-free energy, while $\beta=\gamma T_{C}^{\mu_{\psi}-1}$ results
from integrating the conformable dynamic evolution equation, which
accounts for non-equilibrium scaling governed by the conformable derivative.
They agree only when the deformation vanishes.

\begin{figure*}
\includegraphics[width=0.72\textwidth]{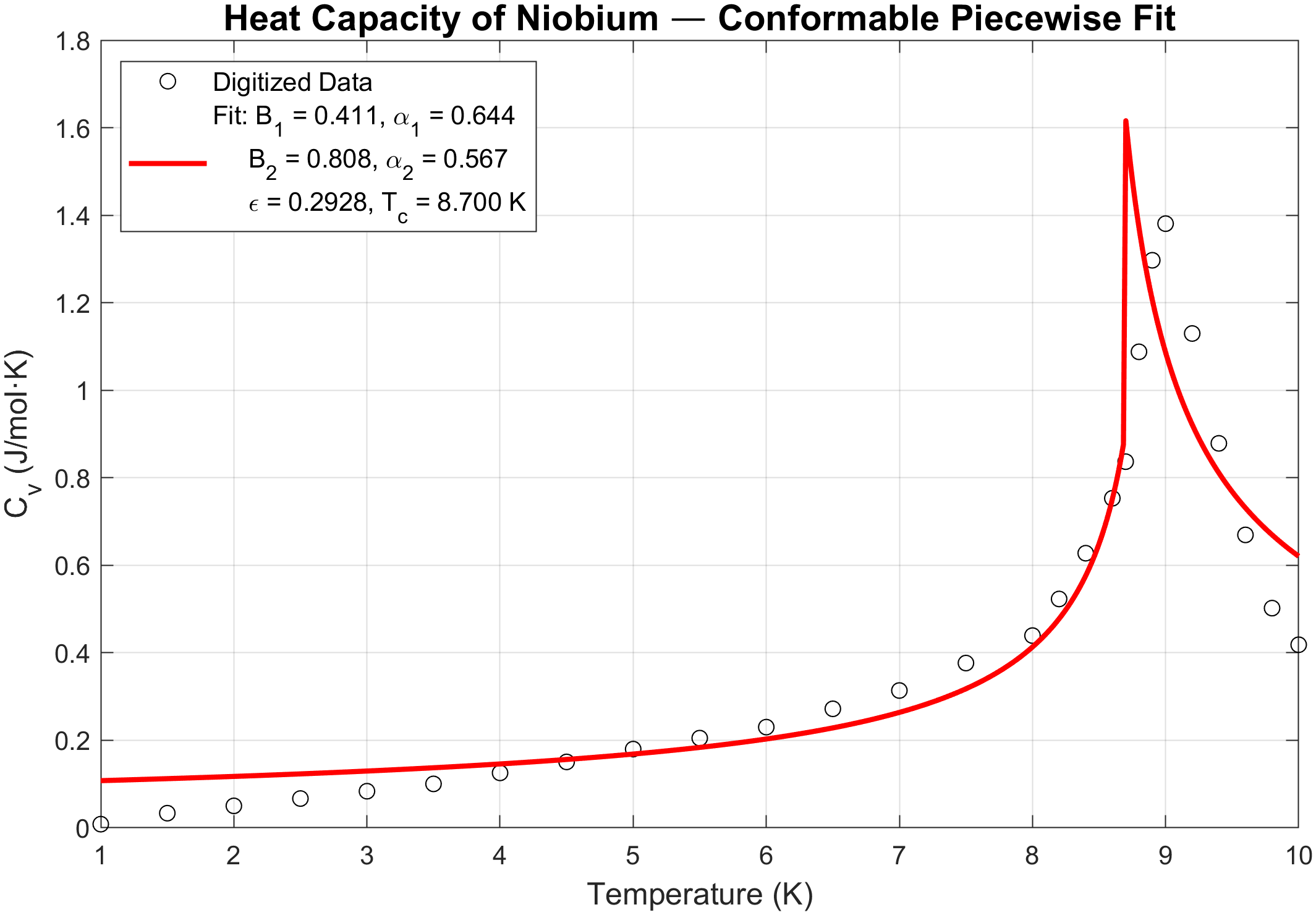}
\caption{Heat capacity of niobium fitted by the smoothed conformable piecewise
model (\ref{eq:Cv_regularized}) with $C_{V}(T)=B_{i}(|T-T_{c}|+\epsilon)^{-\alpha_{i}}$.
Fitted parameters with 95\% confidence intervals: $B_{1}=0.411\pm0.023$,
$\alpha_{1}=0.644\pm0.031$ ($T<T_{c}$); and $B_{2}=0.808\pm0.041$,
$\alpha_{2}=0.567\pm0.028$ ($T>T_{c}$). We used the smoothing factor
$\epsilon=0.293\pm0.015$ to ensure regularization near the divergence
at the critical temperature $T_{c}=8.700\pm0.005$ K. Data from \cite{brown1953}.
The fitting employed weighted least-squares with weights proportional
to $1/\sigma_{i}^{2}$ where $\sigma_{i}$ are experimental uncertainties.}
\label{fig:niobium_Cv_fit} 
\end{figure*}

\section{Numerical fitting and data analysis}

\label{sec:numerical}

To test the predictive power of the conformable formalism, we apply
it to experimental data for the superconducting transition of niobium.

\subsection{Heat capacity of niobium superconductor}

Figure~\ref{fig:niobium_Cv_fit} shows the experimental heat capacity
data for niobium near its superconducting transition, along with a
conformable piecewise fit that captures the critical behavior on both
sides of the transition temperature $T_{c}=8.7$~K. The fitted parameters
are provided in the caption. For the fit we used the regularized model
(\ref{eq:Cv_regularized}). While the fit is not perfect, it demonstrates
good agreement with the data over a wider interval around the critical
temperature.

\subsection{Penetration depth and conformable scaling}

The temperature dependence of the London penetration depth $\lambda_{L}(T)$
provides key informations about the superconducting coherence and
the fluctuation behavior near the critical temperature $T_{c}$. Analogous
to the divergence observed in the specific heat, the penetration depth
exhibits a critical-like behavior and can be modeled using the conformable
scaling framework.

We adopt the piecewise model (\ref{eq:lambdaL_piecewise_alpha}) with
the regularized form (\ref{lambdareg}), where $\epsilon>0$ is the
smoothing parameter that ensures a regularized behavior at the transition
point. The conformable derivative framework again justifies this scaling
law via the modified differential relation (\ref{eq:conformable_lambda_eq})
such that $\lambda_{L}(T)\sim|T-T_{c}|^{-\gamma T_{c}^{\mu-1}}$.

\begin{figure*}
\includegraphics[width=0.72\textwidth]{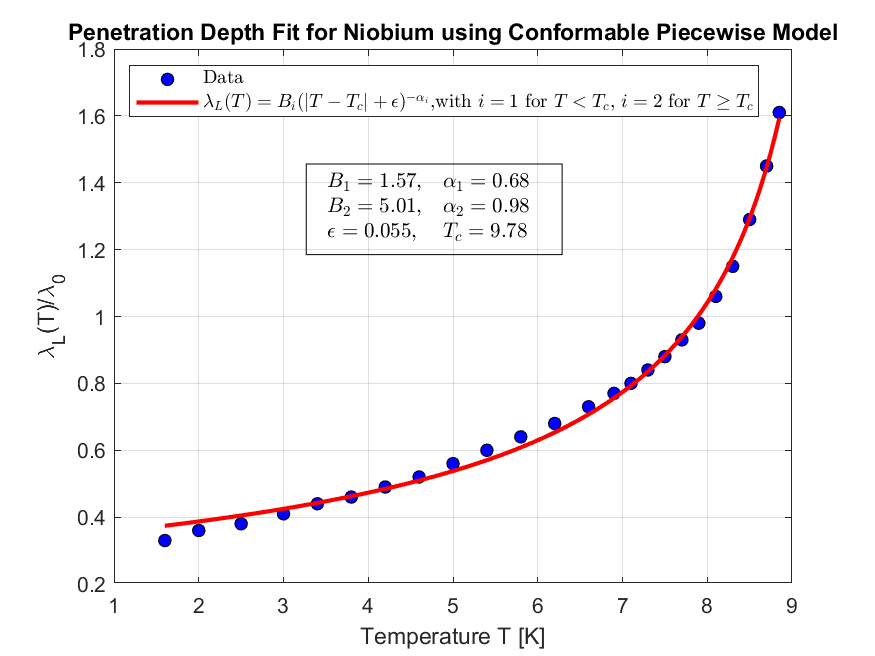}
\caption{Fit of the normalized London penetration depth $\lambda_{L}(T)/\lambda_{0}$,
where $\lambda_{0}$ denotes the London penetration depth extrapolated
to zero temperature, serving as a normalization constant for comparison
between theory and experiment. The data for niobium are fitted using
the conformable piecewise model {[}Eq.~(\ref{eq:lambdaL_piecewise_alpha}){]}
with the regularized form $\lambda_{L}(T)=B_{i}\left(|T-T_{c}|+\epsilon\right)^{-\alpha_{i}},$
as given in Eq.~(\ref{lambdareg}). The fitted parameters are $B_{1}=1.57$,
$\alpha_{1}=0.68$ for $T<T_{c}$ and $B_{2}=5.01$, $\alpha_{2}=0.98$
for $T\protect\geq T_{c}$, with $\epsilon=0.055$ and $T_{c}=9.78~\text{K}$.
Experimental data are taken from \cite{maxfield1965}.}
\label{fig:lambdaL_fit} 
\end{figure*}

The temperature dependence of the normalized London penetration depth
for niobium is shown in Fig.~\ref{fig:lambdaL_fit}. The data are
well described by the conformable piecewise model with distinct scaling
exponents below and above the critical temperature. The fitted parameters
are listed in the caption of Fig.~\ref{fig:lambdaL_fit}.

\subsection{Coherence length from $H_{c2}(T)$ in niobium}

To further validate our theoretical predictions, we turn to experimental
data on niobium and extract the temperature-dependent coherence length
using the upper critical field $H_{c2}(T)$.

\subsubsection{Theoretical background}

In the GL-framework, the upper critical magnetic field $H_{c2}(T)$
is related to the superconducting coherence length $\xi(T)$ by the
expression\textcolor{blue}{{} \cite{Werthamer1966,tinkham1996,DeGennes1999}}
\begin{equation}
H_{c2}(T)=\frac{\Phi_{0}}{2\pi\xi^{2}(T)},\label{eq:Hc2_xi_relation}
\end{equation}
where $\Phi_{0}=h/2e\approx2.07\times10^{-15}$ Wb is the magnetic
flux quantum. This inverse-square relation permits the extraction
of $\xi(T)$ from experimental measurements of $H_{c2}(T)$.

Near the critical temperature $T_{c}$, the coherence length diverges
according to a power law {[}Eq.~(\ref{eq:xi_power_law}){]}. Such
scaling behavior arises naturally when the Ginzburg--Landau framework
is extended to include fractional or conformable spatial derivatives.
These generalized operators account for nonlocality, fractality, and
memory in the order parameter field, yielding a power-law divergence
of the correlation length and critical exponents beyond mean-field
values \cite{Laskin2000FractionalQuantumMechanics,Tarasov2010,Weberszpil2015,weberszpil2016variational,West2020}.
Similar approaches using fractional or stochastic generalizations
of the GL-equation have also been explored in dynamical settings \cite{Zhang2022}.
To see this, consider the modified GL-free energy with the conformable
derivative 
\begin{equation}
|D_{x}^{(\mu)}\psi(x)|^{2}=x^{2(1-\mu)}\left(\frac{d\psi}{dx}\right)^{2},\label{kinetic}
\end{equation}
which replaces the standard gradient term $|\nabla\psi|^{2}$. The
corresponding GL-equation in one dimension becomes 
\begin{eqnarray}
 &  & a_{0}(T-T_{c})\psi+b|\psi|^{2}\psi\nonumber \\
 &  & -\frac{1}{2m}\left[x^{2(1-\mu)}\frac{d^{2}\psi}{dx^{2}}+(1-\mu)x^{1-2\mu}\frac{d\psi}{dx}\right]=0.\label{glone}
\end{eqnarray}

Assuming a linearized form near $T_{c}$ and a trial solution $\psi(x)=\psi_{0}e^{-x/\xi}$,
we compute the derivatives 
\begin{align}
\frac{d\psi}{dx} & =-\frac{\psi_{0}}{\xi}e^{-x/\xi},\\
\frac{d^{2}\psi}{dx^{2}} & =\frac{\psi_{0}}{\xi^{2}}e^{-x/\xi}.
\end{align}
Substituting these into the kinetic term (\ref{kinetic}) produces
\begin{equation}
\text{Kinetic term}\approx\frac{1}{2m}\left[\frac{x^{2(1-\mu)}}{\xi^{2}}+\frac{(1-\mu)x^{1-2\mu}}{\xi}\right]\psi(x).
\end{equation}
Balancing with the linear term $a_{0}(T-T_{c})\psi$ in relation (\ref{glone})
we find 
\begin{equation}
a_{0}(T-T_{c})\sim\frac{1}{2m}\left[\frac{x^{2(1-\mu)}}{\xi^{2}}+\frac{(1-\mu)x^{1-2\mu}}{\xi}\right].
\end{equation}

Let $\tau=T_{c}-T>0$ be the reduced temperature. The modified Ginzburg--Landau
(GL) equation involves two characteristic contributions: the standard
term $T_{1}(\tau,\xi)\propto\xi^{-2}$, associated with the spatial
gradient energy, and the conformable (or fractional) correction $T_{2}(\tau,\xi)\propto\tau$,
which encodes temperature-dependent scaling.

The relative magnitude of these two terms determines the dominant
physical regime. The crossover point is obtained when both contributions
are of comparable strength, 
\begin{equation}
T_{1}(\tau_{*},\xi_{*})=T_{2}(\tau_{*},\xi_{*}),
\end{equation}
which defines a characteristic temperature scale $\tau_{*}$ and a
corresponding coherence length $\xi_{*}$. In practice, $\tau_{*}$
marks the boundary between the mean-field region and the conformable-dominated
region, and $\xi_{*}$ is the coherence length at this crossover.

As $T\to T_{c}$, the reduced temperature becomes very small ($\tau\ll\tau_{*}$)
and the coherence length grows ($\xi\gg\xi_{*}$). In this regime,
the conformable correction dominates, leading to the scaling law 
\begin{equation}
\xi(T)\sim(T_{c}-T)^{-1}.
\end{equation}
In the opposite regime, defined by $\tau\gg\tau_{*}$ (equivalently
$\xi\ll\xi_{*}$), the standard GL term prevails, and the classical
mean-field result is recovered, 
\begin{equation}
\xi(T)\sim(T_{c}-T)^{-1/2}.
\end{equation}
Therefore, $(\tau_{*},\xi_{*})$ act as crossover parameters separating
two asymptotic domains: a critical region governed by conformable
scaling and a mean-field region governed by conventional GL behavior.
Thus, the power-law divergence (\ref{eq:xi_power_law}) is recovered
with an effective exponent $\nu$ that may vary depending on which
term dominates and the value of the conformable parameter $\mu$.
In fractional generalizations of the GL-theory, one may therefore
consider replacing the usual spatial gradient term $|\nabla\psi|^{2}$
by this deformed expression involving the conformable derivative,
leading to a modified scaling law for $\xi(T)$.

The experimental data for $H_{c2}(T)$ were extracted from Fig.~4
of the classical work by Williamson \cite{williamson1970bulk}, which
presents the behavior of the upper critical field for niobium. The
experimental points (open circles) were digitized using the \textit{WebPlotDigitizer}
tool.

The original figure displays the data in a dimensionless form, with
the vertical axis representing the ratio $-H_{c2}^{2}/\left(dH_{c2}/dt\right)_{t=1}$
as a function of the reduced temperature $t=T/T_{c}$. To reconstruct
the actual values of $H_{c2}(T)$ in physical units (Tesla), we assumed
a functional dependence near the critical transition of the form 
\begin{equation}
H_{c2}(T)=H_{c2}(0)\left(1-\frac{T}{T_{c}}\right)^{n},\label{eq:Hc2_model}
\end{equation}
where $H_{c2}(0)$ denotes the extrapolated upper critical field at
zero temperature and $n$ is a fitting exponent. The critical temperature
was fixed at $T_{c}=9.2$ K, consistent with values for pure niobium
and in line with the parameters adopted in literature \cite{tinkham1996}.
After conversion, the data were fitted with the empirical model (\ref{eq:Hc2_model}),
enabling the subsequent extraction of the temperature dependence of
the coherence length $\xi(T)$ through inversion of relation (\ref{eq:Hc2_xi_relation}),
i.e., 
\begin{equation}
\xi(T)=\sqrt{\frac{\Phi_{0}}{2\pi H_{c2}(T)}}.\label{eq:xi_from_Hc2}
\end{equation}
This procedure provides an experimental foundation to assess the validity
of theoretical models for the temperature dependence of the coherence
length.

We note that here we claim that the use of a conformable derivative
in the GL-framework is motivated by the need to capture mesoscopic
and spatially inhomogeneous effects that are not adequately described
by the standard theory. In particular, real superconducting samples
often exhibit intrinsic disorder, nonuniform pinning landscapes, or
microstructural granularity, leading to effective nonlocal interactions
and spatially varying coherence. The conformable derivative introduces
a scale-dependent deformation of the spatial gradient, which can mimic
the impact of fractal-like geometries or finite-size scaling in constrained
domains. This modification allows for a more flexible description
of the kinetic term and leads to a generalized coherence length scaling,
making the theory more compatible with experimental deviations from
classical mean-field behavior.

\subsubsection{Results and analysis}

\begin{figure*}
\includegraphics[width=0.72\textwidth]{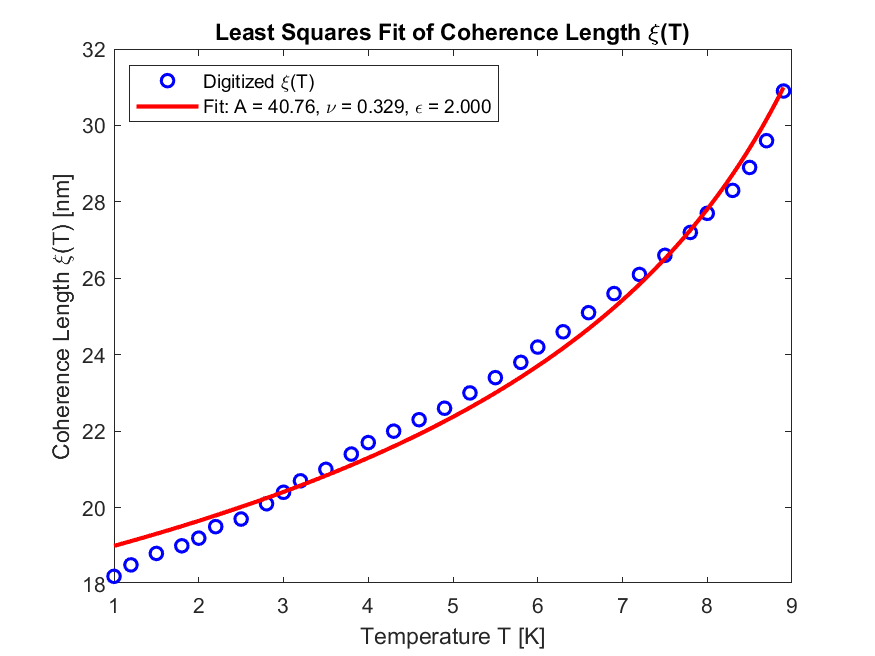}
\caption{Extracted coherence length $\xi(T)$ (blue dots) and fitted model
(red line) using $\xi(T)=A(T_{c}-T)^{-\nu}$ with $\nu=0.329$ and
$A=40.76$. Experimental data adapted from \cite{williamson1970bulk},
see text.}
\label{fig:xi_fit} 
\end{figure*}

Figure~\ref{fig:xi_fit} displays the digitized data and the fitted
model. The excellent agreement indicates the suitability of the power-law
model in this temperature range. The coherence length $\xi(T)$ describes
the spatial extent over which the superconducting order parameter
$\psi$ varies appreciably. As $T\to T_{c}$, $\xi(T)\to\infty$,
and $A$ sets the overall magnitude of $\xi$ for temperatures below
$T_{c}$. This coefficient $A$ encodes material-specific properties,
including microscopic interactions, effective mass of Cooper pairs,
and strength of the pairing potential.

The extracted critical exponent $\nu\approx0.329$ is notably smaller
than the classical GL-mean-field prediction of $\nu=1/2$. Several
factors may contribute to this deviation: (i) A limited resolution
or experimental uncertainty in the high-temperature region ($T\to T_{c}$);
(ii) residual disorder or impurity-induced scattering effects not
incorporated into the ideal GL-framework; (iii) memory effects---either
thermal or spatial---captured through conformable (fractional-order)
derivatives; (iv) effective averaging over anisotropic coherence lengths
in polycrystalline or textured samples.

From the expression 
\begin{equation}
\xi(T)=A(T_{c}-T)^{-\nu},
\end{equation}
we deduce the units of the coefficient $A$ by dimensional analysis.
Since $\xi(T)$ is given in nanometers (nm), $(T_{c}-T)$ in Kelvin
(K), and $\nu$ is dimensionless, $A$ must have units of $\text{nm}\cdot\text{K}^{\nu}$.
Given the fitted exponent $\nu\approx0.329$ and $A=40.76$, this
means that 
\begin{equation}
\xi(T)\approx40.76(T_{c}-T)^{-0.329}\quad\text{[in nm]}.
\end{equation}
To express $A$ in SI units (meters), we convert according to 
\begin{equation}
A_{\text{SI}}=40.76\times10^{-9}\ \text{m}\cdot\text{K}^{0.329}=4.076\times10^{-8}\ \text{m}\cdot\text{K}^{0.329}.
\end{equation}
Such a reduced exponent is qualitatively compatible with generalized
GL-models that incorporate non-integer-order differential structures.

For clean elemental superconductors like niobium, coherence lengths
on the order of 30-40 nm are typical at intermediate temperatures.
The value $A=40.76nm\cdot\text{K}^{\nu}$ is thus physically reasonable
and consistent with known properties of niobium. It represents the
critical amplitude of the coherence length near the critical temperature
$T_{c}$. Its physical meaning and units depend on how the data was
scaled in the fitting process. A larger $A$ implies that $\xi(T)$
becomes large faster as $T\to T_{c}$, for a fixed $\nu$. This can
be influenced by material purity, anisotropy or also electron-phonon
coupling strength.

\section{Conclusions and outlook}

In this work, we developed and applied the conformable derivative
framework to describe critical phenomena near continuous phase transitions.
By deforming the standard derivative operator through a temperature-
or space-dependent scaling, we derived modified evolution equations
for key thermodynamic observables, including the heat capacity, magnetization,
susceptibility, and correlation length. These equations yield consistent
power-law behaviors near the critical point, with critical exponents
expressed analytically in terms of conformable parameters. The formalism
preserves dimensional consistency and remains compatible with equilibrium
thermodynamics, while offering greater flexibility than standard approaches.
In the limiting case $\mu\to1$, the conformable derivative reduces
to the standard derivative, and the classical Ginzburg-Landau theory
with mean-field critical exponents is recovered.

In particular, the observed deviations from the mean-field exponents
are explained as consequences of conformable deformation, capturing
finite-size effects, spatial granularity, and critical slowing down.
The flexibility of the deformation parameter enables the modeling
of crossover behaviors between distinct universality classes, potentially
capturing nontrivial fixed-point trajectories. The conformable approach
thus provides a meaningful interpolation between classical mean-field
theory and more general statistical frameworks, such as nonextensive
thermodynamics.

We demonstrated the relevance of the conformal framework by fitting
of experimental data of the superconducting transition of niobium.
Good fits could be achieved over relatively broad intervals around
the critical temperature for the heat capacity, the London penetration
depth, and the coherence length. For this purpose, an asymmetric formulation
was proposed based on the conformal GL-formulation, in which different
model parameters appear below and above the critical temperature.

Contrary to the impression that the present formulation merely restates
known mean-field scaling, our results highlight several nontrivial
\emph{novel physical insights}:

\textit{(i) Dynamical origin of static exponents.} By combining the
conformable derivative with the critical kinetic coefficient $\Gamma(T)\sim|T-T_{c}|^{z\nu}$,
the static exponents emerge directly from the relaxation dynamics,
rather than being postulated. For instance, the order parameter equation
yields the form 
\[
\Gamma(T)T^{1-\mu_{\psi}}\frac{d\psi}{dT}\simeq-a_{0}(T_{c}-T)\psi,
\]
leading to $\psi(T)\propto(T_{c}-T)^{\beta(\mu_{\psi})}$, where $\beta$
is determined by $(\mu_{\psi},\Gamma)$. This mechanism is absent
in a simple reparametrizations of $T$.

\textit{(ii) Cross-observable constraints.} Equation~(\ref{eq:CrossScaling})
establishes the internal consistency among the conformable parameters
associated with different observables. Specifically, the exponents
governing the temperature dependence of $\{C_{V},|M|,\chi,\xi\}$
are not independent but related by the classical scaling identities
expressed in terms of the deformation indices $\mu_{X}$: 
\begin{equation}
\mu_{C}^{-1}=2\,\mu_{M}^{-1}+\mu_{\chi}^{-1}-2\,\mu_{\xi}^{-1}.\label{eq:CrossScaling}
\end{equation}
Hence, once $\mu_{C}$ is determined experimentally, the remaining
parameters $\{\mu_{M},\mu_{\chi},\mu_{\xi}\}$ are constrained by
this relation. This feature defines a set of cross-observable tests
for the conformable scaling hypothesis, providing a stringent criterion
for internal consistency across thermodynamic observables. 

\textit{(iii) Distinct experimental signatures.} The prediction that
$T^{\mu_{X}}d\ln X/dT$ has a constant slope in $\ln|T-T_{c}|$ provides
a direct discriminant from conventional fits: the slope is observable-dependent
and determined by microscopic couplings, not just by a universal change
of variables.

\textit{(iv) Physical interpretation of $\mu$.} The conformable index
$\mu$ captures finite-size effects, disorder, and nontrivial memory
of thermal fluctuations, bridging kinetic slowing down and effective
fractal dimensionality. This extends beyond a notational reformulation.

A significant advantage of the conformable framework lies in its analytical
tractability. Unlike renormalization group methods, which often rely
on asymptotic expansions and perturbative schemes, the conformable
model yields closed-form expressions that can be directly fitted to
experimental data. This was demonstrated through application to superconducting
phase transitions in niobium, where the model successfully captured
the asymmetric scaling of specific heat and London penetration depth
near the critical temperature. The extracted exponents are consistent
with deviations from mean-field theory and suggest the presence of
mesoscopic effects, spatial inhomogeneities, and memory-driven dynamics.

The geometric interpretation of the conformable framework developed
herein is linked to deformations in the thermal or spatial metric
\cite{Weberszpil2015,rosa2018dual,HasYilmazBaleanu2024} and offers
insight into how fractal geometry and nonlocal responses might influence
critical behavior. In parallel, conformable symmetry approaches have
also found applications in nuclear structure and critical point symmetries
\cite{RaissiRad2021}, illustrating the broader relevance of such
deformations across diverse physical systems. In particular, the framework
is general and can be adapted to a broad range of critical systems,
including percolation thresholds, spin glasses, disordered media,
and quantum phase transitions. Its generality makes it a valuable
tool for exploring universality in complex systems, particularly in
non-equilibrium or nonextensive regimes.

Beyond phenomenological applications, several open questions remain
as to the rigorous mathematical formulation of the conformable framework.
These include the development of a well-defined operator semigroup
structure, the spectral theory of deformed differential operators,
and the formulation of variational principles compatible with conformable
dynamics \cite{godinho2020variational}. Addressing these foundational
issues may reveal deeper connections between the conformable approach
and the broader structure of mathematical physics.

The possible extension of the conformal framework to quantum phase
transitions, non-equilibrium critical dynamics, and systems with complex
topology will be discussed in the future. These applications require
further theoretical development and experimental validation.
\begin{acknowledgments}
JW wishes to express their gratitude to FAPERJ, APQ1, for partial
financial support. RM acknowledges the German Science Foundation (DFG,
grant ME 1535/22-1) for support.
\end{acknowledgments}

\section*{Declarations}

\subsection*{Conflicts of interest/Competing interests:}

The authors declare no conflicts of interest related to this work.

\subsection*{Declaration of generative AI and AI-assisted technologies in the
writing process}

During the preparation of this work the author(s) used ChatGpt in
order to impriove the english. After using this tool/service, the
author(s) reviewed and edited the content as needed and take(s) full
responsibility for the content of the publication.

\appendix

\section{Dimensional consistency}

The mathematical structure of the conformable derivative suggests
a deeper geometric meaning. In this section, we explore interpretations
based on fractal metrics and spatial deformation. Specifically, we
analyze the dimensional consistency of the primary equations and critical
exponents derived using the conformable derivative framework.

\subsection{Conformable derivative units}

The conformable derivative with respect to temperature is defined
as 
\begin{equation}
D_{T}^{(\mu)}f(T)=T^{1-\mu}\frac{df}{dT}
\end{equation}
Given that temperature has units of Kelvin, $[T]=\text{K}$ we see
that $\frac{df}{dT}$ has units of $[f]/\text{K}$. Therefore 
\begin{equation}
[D_{T}^{(\mu)}f]=[f]\cdot\mathrm{K}^{-\mu}.
\end{equation}
This confirms that the conformable derivative scales as expected,
preserving consistent units when applied to temperature-dependent
quantities.

\subsection{Critical exponent dimensionality}

The critical exponents $\alpha$, $\beta$, $\gamma$, and $\nu$
are defined through 
\begin{eqnarray}
\alpha=\kappa T_{c}^{\mu_{C}-1},\quad\beta=\gamma T_{c}^{\mu_{M}-1},\nonumber \\
\gamma=\lambda T_{c}^{\mu_{\chi}-1},\quad\nu=\rho T_{c}^{\mu_{\xi}-1}.
\end{eqnarray}
Since exponents such as $\alpha$ appear in power-law expressions
of the form $C(T)\sim|T-T_{c}|^{-\alpha}$, they must be dimensionless.
For this to hold, $\kappa$ must have units of $\text{K}^{1-\mu_{C}}$,
and similarly for the other exponents. Hence, with this choice all
critical exponents are indeed dimensionless, satisfying the necessary
conditions for physical consistency.

\subsection{Modified Ginzburg--Landau equation}

The modified GL-equation derived using the conformable kinetic term
reads 
\begin{equation}
a(T)\psi+b|\psi|^{2}\psi-\frac{1}{2m}T^{2(1-\alpha)}\frac{d^{2}\psi}{dT^{2}}-\frac{1-\alpha}{m}T^{1-2\alpha}\frac{d\psi}{dT}=0.
\end{equation}
Now let $\psi$ have the units $[\psi]=\text{K}^{-1/2}$. Then (i)
$a(T)=a_{0}(T-T_{c})$ implies $[a_{0}]=\text{K}^{-2}$; (ii) to match
units, the coefficient $b$ must have the units $[b]=\text{K}^{-5/2}$;
(iii) the kinetic terms involve second and first derivatives, and
with the conformable powers of $T$, they retain the same dimension
as $\psi/\text{K}$, ensuring consistency. Thus, all terms in the
equation have matching units.

\subsection{Scaling of the heat capacity}

The heat capacity scales as 
\begin{equation}
C(T)\sim|T-T_{c}|^{-\alpha},\quad[C]=\frac{E}{T}
\end{equation}
For dimensional consistency, $\alpha$ must be dimensionless, which
is satisfied given that $\alpha=\kappa T_{c}^{\mu_{C}-1}$.

\subsection{Penetration depth and order parameter}

The London penetration depth scales as 
\begin{equation}
\lambda_{L}(T)\sim(T_{c}-T)^{-\zeta},\quad\zeta=\beta/2.
\end{equation}
Assuming that $[\lambda_{L}]=\text{length}$, and that $\zeta$ is
dimensionless, the units are correctly preserved.

In summary, all core expressions and critical exponent definitions
derived from the conformable derivative framework are dimensionally
consistent. This ensures the robustness of the theoretical framework
and its applicability to thermodynamical systems such as superconducting
transitions.

\section{Variational derivation of the modified Ginzburg-Landau equation}

\label{sec:Variational-Derivation-of}

We start from the conformable free energy functional 
\begin{equation}
\mathcal{F}_{\alpha}[\psi]=\int\left[a(T)|\psi|^{2}+\frac{b}{2}|\psi|^{4}+\frac{1}{2m}\left|T_{\alpha}\psi(T)\right|^{2}\right]dT.
\end{equation}
The conformable derivative with respect to temperature reads 
\begin{equation}
T_{\alpha}\psi(T)=T^{1-\alpha}\frac{d\psi}{dT}.
\end{equation}
Then, the kinetic term takes on the form 
\begin{equation}
\left|T_{\alpha}\psi(T)\right|^{2}=T^{2(1-\alpha)}\left|\frac{d\psi}{dT}\right|^{2}
\end{equation}
and the full functional reads 
\begin{equation}
\mathcal{F}_{\alpha}[\psi]=\int\left[a(T)|\psi|^{2}+\frac{b}{2}|\psi|^{4}+\frac{1}{2m}T^{2(1-\alpha)}\left|\frac{d\psi}{dT}\right|^{2}\right]dT.
\end{equation}

To minimize the free energy, we compute the variation with respect
to $\psi^{*}(T)$, 
\begin{eqnarray}
\delta\mathcal{F}_{\alpha} & = & \int\left[a(T)\delta|\psi|^{2}+b|\psi|^{2}\delta|\psi|^{2}\right.\nonumber \\
 &  & \hspace*{-1.6cm}\left.+\frac{1}{2m}T^{2(1-\alpha)}\left(\frac{d\psi}{dT}\frac{d(\delta\psi^{*})}{dT}+\frac{d\psi^{*}}{dT}\frac{d(\delta\psi)}{dT}\right)\right]dT.
\end{eqnarray}
Using integration by parts and assuming vanishing boundary terms,
\begin{eqnarray}
 &  & \int T^{2(1-\alpha)}\frac{d\psi}{dT}\frac{d(\delta\psi^{*})}{dT}dT\nonumber \\
 &  & =-\int\delta\psi^{*}\left[\frac{d}{dT}\left(T^{2(1-\alpha)}\frac{d\psi}{dT}\right)\right]dT.
\end{eqnarray}
Therefore, setting $\delta\mathcal{F}_{\alpha}=0$, the associated
Euler--Lagrange equation becomes 
\begin{equation}
a(T)\psi+b|\psi|^{2}\psi-\frac{1}{2m}\frac{d}{dT}\left(T^{2(1-\alpha)}\frac{d\psi}{dT}\right)=0.
\end{equation}
Expanding the total derivative, we obtain 
\begin{eqnarray}
 &  & a(T)\psi+b|\psi|^{2}\psi-\frac{1}{2m}T^{2(1-\alpha)}\frac{d^{2}\psi}{dT^{2}}\nonumber \\
 &  & \hspace*{0.8cm}-\frac{(1-\alpha)}{m}T^{1-2\alpha}\frac{d\psi}{dT}=0.\label{eulerlag}
\end{eqnarray}

\subsection*{Final form of in powers of $1-T/T_{c}$}

Substituting the ansatz 
\begin{equation}
\psi(T)=\psi_{0}\left(1-\frac{T}{T_{c}}\right)^{\beta}\label{eq:ansatz}
\end{equation}
into the full equation, we obtain 
\begin{eqnarray}
 &  & a_{0}(T-T_{c})\psi_{0}\left(1-\frac{T}{T_{c}}\right)^{\beta}+b\psi_{0}^{3}\left(1-\frac{T}{T_{c}}\right)^{3\beta}\nonumber \\
 &  & \quad-\frac{1}{2m}T^{2(1-\alpha)}\frac{\beta(\beta-1)\psi_{0}}{T_{c}^{2}}\left(1-\frac{T}{T_{c}}\right)^{\beta-2}\nonumber \\
 &  & \quad+\frac{(1-\alpha)}{m}T^{1-2\alpha}\frac{\beta\psi_{0}}{T_{c}}\left(1-\frac{T}{T_{c}}\right)^{\beta-1}=0.
\end{eqnarray}
Matching powers of $(1-T/T_{c})$, we recover $\beta=1/2$ as the
critical exponent.

Now, we analyze the behavior of $\psi(T)$ near the critical temperature
$T_{c}$. To this end, consider again the nonlinear temperature-deformed
GL-type equation 
\begin{eqnarray}
a(T)\psi+b|\psi|^{2}\psi-\frac{1}{2m}T^{2(1-\alpha)}\frac{d^{2}\psi}{dT^{2}}\nonumber \\
-\frac{1-\alpha}{m}T^{1-2\alpha}\frac{d\psi}{dT}=0,
\end{eqnarray}
with $a(T)=a_{0}(T-T_{c})$. We again assume the ansatz 
\begin{equation}
\psi(T)=\psi_{0}\left(1-\frac{T}{T_{c}}\right)^{\beta}=\psi_{0}\varepsilon^{\beta},\quad\text{where }\varepsilon=1-\frac{T}{T_{c}}.
\end{equation}
Then, forming the derivatives with respect to T, we have 
\begin{eqnarray}
 &  & \frac{d\varepsilon}{dT}=-\frac{1}{T_{c}},\nonumber \\
 &  & \frac{d\psi}{dT}=-\frac{\beta\psi_{0}}{T_{c}}\varepsilon^{\beta-1}\nonumber \\
 &  & \frac{d^{2}\psi}{dT^{2}}=\frac{\beta(\beta-1)\psi_{0}}{T_{c}^{2}}\varepsilon^{\beta-2}.
\end{eqnarray}

We then obtain the different terms. Starting with the linear term,
\begin{equation}
a_{0}(T-T_{c})\psi=-a_{0}T_{c}\psi_{0}\varepsilon^{\beta+1};
\end{equation}
then the nonlinear term 
\begin{equation}
b|\psi|^{2}\psi=b\psi_{0}^{3}\varepsilon^{3\beta};
\end{equation}
the term with the second derivative, 
\begin{equation}
-\frac{1}{2m}T^{2(1-\alpha)}\frac{\beta(\beta-1)\psi_{0}}{T_{c}^{2}}\varepsilon^{\beta-2};
\end{equation}
and the term with the first-order derivative, 
\begin{equation}
\frac{(1-\alpha)\beta\psi_{0}}{mT_{c}}T^{1-2\alpha}\varepsilon^{\beta-1}.
\end{equation}

We see that $\varepsilon=1-T/T_{c}$ enters these different terms
with different exponent. Seeking a consistent balance, we suppose
that the nonlinear term $\propto\varepsilon^{3\beta}$ and the linear
term $\propto\varepsilon^{\beta+1}$ dominate. We then equate their
powers: $3\beta=\beta+1$, implying $\beta=1/2$. This is consistent
with standard Landau theory and a typical critical exponent in second-order
phase transitions. This means that in the four terms above, we have
the orders 
\begin{equation}
\begin{aligned} & \varepsilon^{\beta-2}=\varepsilon^{-3/2}\\
 & \varepsilon^{\beta-1}=\varepsilon^{-1/2}\\
 & \varepsilon^{3\beta}=\varepsilon^{3/2}\\
 & \varepsilon^{\beta+1}=\varepsilon^{3/2}
\end{aligned}
,
\end{equation}
i.e., the derivative terms diverge as $\varepsilon\to0$. However,
despite their divergence, the derivative terms describe kinetic effects
(gradients of $\psi$). Near $T=T_{c}$, the order parameter becomes
small and smooth, $\psi\to0$, and its variation is slow. Finally
the mean-field and equilibrium theories focus on minimizing the free
energy---gradient terms are subleading. This allows us to match 
\begin{equation}
a(T)\psi+b\psi^{3}=0\quad\Rightarrow\quad\psi=\pm\sqrt{-\frac{a(T)}{b}}.
\end{equation}
From this we conclude that 
\begin{equation}
\psi(T)=\psi_{0}\left(1-\frac{T}{T_{c}}\right)^{1/2}
\end{equation}
Thus, the derivative terms represent corrections to the scaling behavior,
but not to the leading-order critical exponent. This demonstrates
that the ansatz (\ref{eq:ansatz}) satisfies the nonlinear equation
near $T=T_{c}$, with leading order balance provided by the algebraic
terms, 
\begin{equation}
a(T)\psi+b\psi^{3}=0\quad\Rightarrow\quad\boxed{\beta=1/2}.
\end{equation}

This matches classical Landau theory and remains valid under conformable
kinetic corrections. Moreover, this represents a static equilibrium
solution for which the derivatives are zero (i.e., long-time equilibrium).
It is important to note that our model for equilibrium and homogeneous
conditions and when fluctuations are negligible, has the classical
exponent $\beta=1/2$.

\section{Recovery of the relaxation equation in the undeformed limit}

We now show that the classical critical exponent $\beta=\tfrac{1}{2}$
naturally emerges from the conformable relaxation equation in the
undeformed equilibrium limit. Consider the generalized relaxation
model 
\begin{equation}
\Gamma(T)\,T^{1-\mu_{\psi}}\frac{d\psi}{dT}=-a_{0}(T_{c}-T)\psi-b|\psi|^{2}\psi,\label{eq:relaxation_general}
\end{equation}
where $\Gamma(T)$ is a temperature-dependent kinetic coefficient,
$\mu_{\psi}\in(0,1]$ is the conformable deformation parameter, and
the right-hand side arises from the variational derivative of the
standard GL-free energy.

In the undeformed limit, we take 
\begin{equation}
\mu_{\psi}=1,\quad\text{so that}\quad T^{1-\mu_{\psi}}=1,
\end{equation}
and for simplicity we assume that the kinetic coefficient is constant
near $T_{c}$, i.e., $\Gamma(T)\approx\Gamma_{0}$. Equation~(\ref{eq:relaxation_general})
then becomes 
\begin{equation}
\Gamma_{0}\frac{d\psi}{dT}=-a_{0}(T_{c}-T)\psi-b|\psi|^{2}\psi.\label{eq:undeformed_relaxation}
\end{equation}
We now look for a solution near the critical point of the form 
\begin{equation}
\psi(T)=\psi_{0}(T_{c}-T)^{\beta},
\end{equation}
valid for $T<T_{c}$ and where $\psi_{0}$ is a constant amplitude
to be determined. Computing the derivative, we have 
\begin{equation}
\frac{d\psi}{dT}=-\beta\psi_{0}(T_{c}-T)^{\beta-1}.
\end{equation}
Substituting into Eq.~(\ref{eq:undeformed_relaxation}), we obtain
\begin{eqnarray}
 &  & -\Gamma_{0}\beta\psi_{0}(T_{c}-T)^{\beta-1}\nonumber \\
 &  & =-a_{0}(T_{c}-T)\psi_{0}(T_{c}-T)^{\beta}-b\psi_{0}^{3}(T_{c}-T)^{3\beta}.
\end{eqnarray}
Divide both sides by $\psi_{0}(T_{c}-T)^{\beta-1}$ (assuming $\psi_{0}\neq0$
and $T<T_{c}$), 
\begin{equation}
-\Gamma_{0}\beta=-a_{0}(T_{c}-T)^{2}-b\psi_{0}^{2}(T_{c}-T)^{2\beta+1}.
\end{equation}
Now we check the leading-order behavior as $T\to T_{c}^{-}$: The
term $(T_{c}-T)^{2}$ dominates, while the nonlinear term $\propto(T_{c}-T)^{2\beta+1}$
is subleading if $2\beta+1>2$ and thus $\beta>\tfrac{1}{2}$; or
they are comparable if $\beta=\tfrac{1}{2}$.

Assuming that the dominant balance occurs between the terms of order
\begin{equation}
-\Gamma_{0}\beta\approx-a_{0}(T_{c}-T)^{2},
\end{equation}
we see that this leads to a contradiction: the left-hand side is constant,
while the right-hand side vanishes as $T\to T_{c}$. Therefore, this
balance fails. Instead, if we assume that the dominant balance occurs
between the first-derivative term and the nonlinear cubic term, 
\begin{equation}
-\Gamma_{0}\beta\psi_{0}(T_{c}-T)^{\beta-1}\approx-b\psi_{0}^{3}(T_{c}-T)^{3\beta},
\end{equation}
and we cancel signs and factors of $\psi_{0}$, we see that 
\begin{equation}
\Gamma_{0}\beta(T_{c}-T)^{\beta-1}=b\psi_{0}^{2}(T_{c}-T)^{3\beta}.
\end{equation}
Divide both sides by $(T_{c}-T)^{\beta-1}$ to find 
\begin{equation}
\Gamma_{0}\beta=b\psi_{0}^{2}(T_{c}-T)^{2\beta+1}.
\end{equation}
To keep both sides finite and nonzero as $T\to T_{c}$, we must have
\begin{equation}
2\beta+1=0\quad\Rightarrow\quad\beta=-\tfrac{1}{2},
\end{equation}
which is unphysical.

So we try instead to match the linear and nonlinear terms (equivalent
to the adiabatic limit), 
\begin{equation}
a_{0}(T_{c}-T)\psi\sim b\psi^{3}\quad\Rightarrow\quad a_{0}(T_{c}-T)\sim b\psi^{2}.
\end{equation}
Solving for $\psi(T)$, we see that 
\begin{equation}
\psi^{2}(T)\sim\frac{a_{0}}{b}(T_{c}-T),\quad\Rightarrow\quad\psi(T)\sim\sqrt{\frac{a_{0}}{b}}(T_{c}-T)^{1/2}.
\end{equation}
Thus, the leading-order balance occurs not with the derivative term,
but with the free-energy terms, confirming that in the long-time (quasi-equilibrium)
limit, the stationary solution satisfies 
\begin{equation}
\beta=\frac{1}{2}.
\end{equation}
This demonstrates that even when starting from the generalized relaxation
equation, the classical mean-field exponent $\beta=1/2$ is recovered
in the undeformed limit $\mu_{\psi}=1$, under equilibrium assumptions.
The derivative terms decay near $T_{c}$, and the critical scaling
is governed by the balance of the potential terms in the GL-free energy.

\subsection*{Higher-order corrections}

Derivative terms can be treated as perturbations (next-to-leading
order). They modify the scaling only slightly and do not affect the
critical exponent to leading order. Thus, (i) $\varepsilon^{\beta-1}$
and $\varepsilon^{\beta-2}$ diverge as $\varepsilon\to0$ but are
derivative-based; (ii) near $T_{c}$, $\psi$ is small and smooth,
so derivatives are small; (iii) the dominant balance is between $a(T)\psi$
and $b\psi^{3}$, yielding again $\beta=1/2$. Thus, indeed the derivative
terms are corrections to this leading-order behavior.

\section{On the claim of simple reparametrization}

For a scalar $f(T)$ with constant coefficients, we have that $D_{T}^{(\mu)}f=T^{1-\mu}f'(T)=df/dT'$
with $T'=T^{\mu}/\mu$. However, consider the two cases:

(i) \textit{Critical kinetics.} With $\Gamma(T)\!\sim\!|T-T_{c}|^{z\nu}$,
the relaxation law becomes $\Gamma(T(T'))\,\tfrac{dX}{dT'}=-\partial F/\partial X$.
The nontrivial $T$-dependence of $\Gamma$ survives, producing a
distinct scaling.

(ii) \textit{Thermal GL-functional.} The conformable GL-free energy
includes a gradient term $\propto T^{2(1-\alpha)}|\tfrac{d\psi}{dT}|^{2}$.
Under a transformation $T\mapsto T'$, Jacobian factors remain, yielding
Euler-Lagrange equations {[}Eq.~(\ref{eulerlag}){]} that are not
reducible to the standard GL-form.

Hence, the conformable framework only collapses to a trivial change
of variables in the static, constant-coefficient case. In the presence
of critical kinetics and weighted GL-terms, it carries genuine physical
content.

\bibliographystyle{apsrev}
\bibliography{references_merged}

@book{stanley1971,
  author = {H. E. Stanley},
  title = {Introduction to Phase Transitions and Critical Phenomena},
  publisher = {Oxford University Press},
  year = {1971}
}

@book{goldenfeld1992,
  author = {N. Goldenfeld},
  title = {Lectures on Phase Transitions and the Renormalization Group},
  publisher = {Addison-Wesley},
  year = {1992}
}

@article{tsallis1988,
  author = {C. Tsallis},
  title = {Possible generalization of Boltzmann-Gibbs statistics},
  journal = {Journal of Statistical Physics},
  volume = {52},
  pages = {479--487},
  year = {1988}
}

@incollection{lenzi,
  author={L. R. Evangelista and E. K. Lenzi},
  title={Fractional diffusion equations and anomalous diffusion},
  booktitle={Fractional diffusion equations and anomalous diffusion},
  publisher={Cambridge University Press, Cambridge UK},
  year={2018},
}

@article{plastino1993,
  author = {A. R. Plastino and A. Plastino},
  title = {Stellar polytropes and Tsallis' entropy},
  journal = {Physics Letters A},
  volume = {174},
  pages = {384--386},
  year = {1993}
}

@article{Hohenberg1977,
  title={Theory of dynamic critical phenomena},
  author={Hohenberg, Pierre C and Halperin, Bertrand I},
  journal={Reviews of Modern Physics},
  volume={49},
  number={3},
  pages={435--479},
  year={1977},
  publisher={American Physical Society},
  doi={10.1103/RevModPhys.49.435}
}

@article{Fisher1972,
  title={Scaling theory for finite-size effects in the critical region},
  author={Fisher, Michael E and Barber, Michael N},
  journal={Physical Review Letters},
  volume={28},
  number={23},
  pages={1516--1519},
  year={1972},
  publisher={American Physical Society},
  doi={10.1103/PhysRevLett.28.1516}
}

@article{Fisher1967,
  author    = {Fisher, M. E.},
  title     = {The theory of critical point singularities},
  journal   = {Reports on Progress in Physics},
  volume    = {30},
  pages     = {615--730},
  year      = {1967},
  doi       = {10.1088/0034-4885/30/2/306}
}

@book{Privman1990,
  title={Finite Size Scaling and Numerical Simulation of Statistical Systems},
  author={Privman, Vladimir},
  year={1990},
  publisher={World Scientific},
  address={Singapore},
  isbn={981-02-0403-1},
  doi={10.1142/1011}
}

@article{Harris1974,
  title={Effect of random defects on the critical behaviour of {I}sing models},
  author={Harris, A Brooks},
  journal={Journal of Physics C: Solid State Physics},
  volume={7},
  number={9},
  pages={1671--1692},
  year={1974},
  publisher={IOP Publishing},
  doi={10.1088/0022-3719/7/9/009}
}

@article{Wilson1971,
  author       = {K. G. Wilson},
  title        = {Renormalization Group and Critical Phenomena. I. Renormalization Group and the Kadanoff Scaling Picture},
  journal      = {Physical Review B},
  volume       = {4},
  number       = {9},
  pages        = {3174--3183},
  year         = {1971},
  doi          = {10.1103/PhysRevB.4.3174},
  publisher    = {American Physical Society}
}

@incollection{Landau1937Eng,
  author       = {L. D. Landau},
  title        = {On the theory of phase transitions},
  booktitle    = {Collected Papers of L. D. Landau},
  editor       = {D. Ter Haar},
  publisher    = {Pergamon Press},
  year         = {1965},
  pages        = {193--216},
  note         = {Originally published in 1937, Zh. Eksp. Teor. Fiz. 7, 19--32}
}

@incollection{Ginzburg1950Eng,
  author       = {V. L. Ginzburg and L. D. Landau},
  title        = {On the theory of superconductivity},
  booktitle    = {Men of Physics: L. D. Landau},
  editor       = {D. Ter Haar},
  publisher    = {Pergamon Press},
  year         = {1965},
  pages        = {546--568},
  note         = {Originally published in 1950, Zh. Eksp Teor. Fiz. 20, 1064--1082}
}

@article{Wilson1974,
  author       = {Kenneth G. Wilson and John Kogut},
  title        = {The Renormalization Group and the {\varepsilon} Expansion},
  journal      = {Physics Reports},
  volume       = {12},
  number       = {2},
  pages        = {75--199},
  year         = {1974},
  doi          = {10.1016/0370-1573(74)90023-4},
  publisher    = {Elsevier},
  note         = {Comprehensive review of the renormalization group and critical phenomena}
}

@book{ZinnJustin2021,
  author    = {Jean Zinn-Justin},
  title     = {Quantum Field Theory and Critical Phenomena},
  publisher = {Oxford University Press},
  edition   = {5th},
  year      = {2021},
  series    = {International Series of Monographs on Physics},
  volume    = {171},
  isbn      = {978-0198834625},
  note      = {Updated and expanded edition with modern developments},
}

@article{Khalil2014,
  author  = {Rabah Khalil and Mohamad Al Horani and Abdelrahim Yousef and Mohammad Sababheh},
  title   = {A New Definition of Fractional Derivative},
  journal = {Journal of Computational and Applied Mathematics},
  volume  = {264},
  pages   = {65--70},
  year    = {2014},
  doi     = {10.1016/j.cam.2014.01.002},
  publisher = {Elsevier}
}

@article{Abdeljawad2015,
  author  = {Thabet Abdeljawad},
  title   = {On conformable fractional calculus},
  journal = {Journal of Computational and Applied Mathematics},
  volume  = {279},
  pages   = {57--66},
  year    = {2015},
  doi     = {10.1016/j.cam.2014.10.016},
  publisher = {Elsevier}
}

@book{Anderson2016,
  author    = {P. W. Anderson},
  title     = {Basic Notions of Condensed Matter Physics},
  publisher = {Westview Press},
  year      = {2016},
  edition   = {2nd},
  isbn      = {978-0813346312},
  note      = {Updated edition of the classic work originally published in 1984}
}

@article{Weberszpil2015,
  author    = {Jos{\'e} Weberszpil and J. A. Helay{\"e}l-Neto},
  title     = {On a connection between a class of q-deformed algebras and the Hausdorff derivative in a medium with fractal metric},
  journal   = {Physica A: Statistical Mechanics and its Applications},
  volume    = {436},
  pages     = {399--404},
  year      = {2015},
  doi       = {10.1016/j.physa.2015.05.002},
  publisher = {Elsevier}
}

@book{tinkham1996,
  title={Introduction to Superconductivity},
  author={Tinkham, Michael},
  edition={2nd},
  year={1996},
  publisher={McGraw-Hill},
  address={New York}
}

@article{williamson1970bulk,
  title={Bulk upper critical field of clean type-II superconductors: V and Nb},
  author={Williamson, SJ},
  journal={Physical review B},
  volume={2},
  number={9},
  pages={3545},
  year={1970},
  publisher={APS}
}

@book{Tinkham2004,
  title={Introduction to Superconductivity},
  author={Tinkham, Michael},
  year={2004},
  publisher={Dover Publications},
  edition={2nd}
}

@article{Bardeen1957,
  title={Theory of superconductivity},
  author={Bardeen, John and Cooper, Leon N and Schrieffer, J Robert},
  journal={Physical Review},
  volume={108},
  number={5},
  pages={1175--1204},
  year={1957}
}

@article{Gorter1934,
  title={On the transition of superconductors to the normal state},
  author={Gorter, C J and Casimir, H B G},
  journal={Physica},
  volume={1},
  number={4},
  pages={306--320},
  year={1934}
}

@article{Park2003,
  title={Strong thermal fluctuations in superconducting films and nanowires},
  author={Park, Ki and Dorsey, Alan T},
  journal={Physical Review Letters},
  volume={91},
  number={15},
  pages={157003},
  year={2003}
}

@book{deGennes1966,
  title={Superconductivity of metals and alloys},
  author={de Gennes, Pierre-Gilles},
  year={1966},
  publisher={W. A. Benjamin}
}

@book{deGennes1972,
  title={Scaling concepts in polymer physics},
  author={de Gennes, Pierre-Gilles},
  year={1979},
  publisher={Cornell University Press}
}

@article{weberszpil2016variational,
	title={Variational approach and deformed derivatives},
	author={Weberszpil, Jos{\'e} and Helay{\"e}l-Neto, Jos{\'e} Abdalla},
	journal={Physica A: Statistical Mechanics and its Applications},
	volume={450},
	pages={217--227},
	year={2016},
	publisher={Elsevier}
}

@article{rosa2018dual,
	title={Dual conformable derivative: Definition, simple properties and perspectives for applications},
	author={Rosa, Wanderson and Weberszpil, Jos{\'e}},
	journal={Chaos, Solitons \& Fractals},
	volume={117},
	pages={137--141},
	year={2018},
	publisher={Elsevier}
}

@article{godinho2020variational,
	title={Variational procedure for higher-derivative mechanical models in a fractional integral},
	author={Godinho, Cresus F de L and Panza, Nelson and Weberszpil, Jos{\'e} and Helay{\"e}l-Neto, JA},
	journal={Europhysics Letters},
	volume={129},
	number={6},
	pages={60001},
	year={2020},
	publisher={IOP Publishing}
}

@article{Sotolongo2021,
	author = {Sotolongo-Costa, Oscar and Weberszpil, Jos/'e},
	title = {Explicit Time-Dependent Entropy Production Expressions: Fractional and Fractal Pesin Relations},
	journal = {Brazilian Journal of Physics},
	volume = {51},
	pages = {635--643},
	year = {2021},
	doi = {10.1007/s13538-021-00897-3}
}

@article{Weberszpil2017_Tissue,
	author = {Weberszpil, Jos/'e and Sotolongo-Costa, Oscar},
	title = {Structural Derivative Model for Tissue Radiation Response},
	journal = {Journal of Advances in Physics},
	volume = {13},
	pages = {4779--4785},
	year = {2017}
}

@article{weberszpil2017generalized,
title={Generalized Maxwell relations in thermodynamics with metric derivatives},
author={Weberszpil, Jos{\'e} and Chen, Wen},
journal={Entropy},
volume={19},
number={8},
pages={407},
year={2017},
publisher={MDPI}
}

@article{Zhang2022,
  author = {Zhang, C.-Q. and Li, H. and Wang, J.},
  title = {Stochastic dynamics of non-autonomous fractional Ginzburg--Landau equations with multiplicative noise},
  journal = {Discrete and Continuous Dynamical Systems - Series B},
  volume = {27},
  number = {5},
  pages = {1765--1789},
  year = {2022},
  doi = {10.3934/dcdsb.2022028}
}

@article{HasYilmazBaleanu2024,
  author = {Has, A. and Yilmaz, B. and Baleanu, D.},
  title = {On the geometric and physical properties of conformable derivative},
  journal = {Mathematical Sciences and Applications E-Notes},
  volume = {12},
  number = {2},
  pages = {60--70},
  year = {2024}
}

@article{RaissiRad2021,
  author = {Raissi-Rad, M. and Gholami, H. and Jafarizadeh, M.A.},
  title = {Conformable fractional E(5) critical point symmetry in nuclear structure},
  journal = {Physica A: Statistical Mechanics and its Applications},
  volume = {574},
  pages = {126056},
  year = {2021},
  doi = {10.1016/j.physa.2021.126056}
}

@article{brown1953,
  author    = {A. Brown and M. W. Zemansky and H. A. Boorse},
  title     = {The Superconducting and Normal Heat Capacities of Niobium},
  journal   = {Physical Review},
  volume    = {92},
  pages     = {52--60},
  year      = {1953},
  month     = oct,
  doi       = {10.1103/PhysRev.92.52}
}

@article{maxfield1965,
  author    = {B. W. Maxfield and W. L. McLean},
  title     = {Superconducting Penetration Depth of Niobium},
  journal   = {Physical Review},
  volume    = {139},
  number    = {5A},
  pages     = {A1515--A1522},
  year      = {1965},
  doi       = {10.1103/PhysRev.139.A1515}
}

@book{Klafter2011,
  title     = {First Steps in Random Walks: From Tools to Applications},
  author    = {Klafter, J. and Sokolov, I. M.},
  publisher = {Oxford University Press},
  address   = {Oxford},
  year      = {2011}
}

@article{Bray1994,
  title   = {Theory of phase-ordering kinetics},
  author  = {Bray, A. J.},
  journal = {Adv. Phys.},
  volume  = {43},
  pages   = {357--459},
  year    = {1994},
  doi     = {10.1080/00018739400101505}
}

@book{Forster1975,
  author    = {Forster, Dieter},
  title     = {Hydrodynamic Fluctuations, Broken Symmetry, and Correlation Functions},
  publisher = {Addison-Wesley},
  address   = {Reading, MA},
  year      = {1975}
}

@article{Metzler2000,
  title   = {The random walk's guide to anomalous diffusion: a fractional dynamics approach},
  author  = {Metzler, R. and Klafter, J.},
  journal = {Phys. Rep.},
  volume  = {339},
  pages   = {1--77},
  year    = {2000},
  doi     = {10.1016/S0370-1573(00)00070-3}
}

@article{Metzler2004,
  author={R. Metzler and J. Klafter},
  title={The restaurant at the end of the random walk: recent developments in
  fractional dynamics descriptions of anomalous dynamical processes},
  journal={J. Phys. A},
  volume={37},
  pages={R161},
  year={2004},
}

@article{Metzler2014,
  title   = {Anomalous diffusion models and their properties: non-stationarity, non-ergodicity, and ageing at the centenary of single particle tracking},
  author  = {Metzler, R. and Jeon, J.-H. and Cherstvy, A. G. and Barkai, E.},
  journal = {Phys. Chem. Chem. Phys.},
  volume  = {16},
  pages   = {24128--24164},
  year    = {2014},
  doi     = {10.1039/C4CP03465A}
}

@article{Sokolov2012,
  title   = {Models of anomalous diffusion in crowded environments},
  author  = {Sokolov, I. M.},
  journal = {Soft Matter},
  volume  = {8},
  pages   = {9043--9052},
  year    = {2012},
  doi     = {10.1039/C2SM25701G}
}

@article{Binder1981,
  title   = {Finite size scaling analysis of Ising model block distribution functions},
  author  = {Binder, K.},
  journal = {Z. Phys. B},
  volume  = {43},
  pages   = {119--140},
  year    = {1981},
  doi     = {10.1007/BF01293604}
}

@article{GodrecheLuck2000,
  title   = {Response of non-equilibrium systems at criticality: ferromagnetic models in dimension two and above},
  author  = {Godr\`eche, Claude and Luck, Jean-Marc},
  journal = {J. Phys. A: Math. Gen.},
  volume  = {33},
  pages   = {9141--9153},
  year    = {2000},
  doi     = {10.1088/0305-4470/33/50/302}
}

@article{Kubo1966,
  author  = {Kubo, Ryogo},
  title   = {The fluctuation-dissipation theorem},
  journal = {Rep. Prog. Phys.},
  volume  = {29},
  pages   = {255--284},
  year    = {1966},
  doi     = {10.1088/0034-4885/29/1/306}
}

@article{HohenbergHalperin1977,
  author  = {Hohenberg, P. C. and Halperin, B. I.},
  title   = {Theory of Dynamic Critical Phenomena},
  journal = {Rev. Mod. Phys.},
  volume  = {49},
  pages   = {435--479},
  year    = {1977},
  doi     = {10.1103/RevModPhys.49.435}
}

@book{Amit2005,
  title     = {Field Theory, the Renormalization Group, and Critical Phenomena},
  author    = {Amit, Daniel J. and Martin-Mayor, Victor},
  edition   = {3rd},
  publisher = {World Scientific},
  year      = {2005},
  isbn      = {978-981-256-047-2},
  doi       = {10.1142/5675}
}

@book{TsallisBook2009,
  author    = {Tsallis, Constantino},
  title     = {Introduction to Nonextensive Statistical Mechanics: Approaching a Complex World},
  publisher = {Springer},
  address   = {New York},
  year      = {2009},
  isbn      = {978-0-387-85358-1},
  doi       = {10.1007/978-0-387-85359-8}
}

@article{Weberszpil2025a,
  author  = {Weberszpil, Jos{\'e}},
  title   = {Microscopic Origins of Conformable Dynamics: From Disorder to Deformation},
  journal = {Physica A: Statistical Mechanics and its Applications},
  year    = {2025},
  doi     = {10.1016/j.physa.2025.130945},
  note    = {in press}
}

@article{Weberszpil2025qdeformation,
  author    = {Weberszpil, Jos\'e},
  title     = {Functional {q}-deformation of orbital velocity in emergent gravitation: Extended framework and galactic applications},
  journal   = {Annals of Physics},
  year      = {2025},
  doi       = {10.1016/j.aop.2025.170246},
  note      = {in press}
}

@book{Landau1978StatisticalPart1,
  title        = {Statistical Physics, Part 1},
  author       = {Landau, Lev Davidovich and Lifshitz, Evgenii Mikhailovich},
  series       = {Course of Theoretical Physics},
  volume       = {5},
  edition      = {3rd},
  year         = {1978},
  publisher    = {Pergamon Press},
  address      = {Oxford},
  isbn         = {0-08-023039-3}
}

@book{mathai2017introduction,
  title={An Introduction to Fractional Calculus},
  author={Mathai, A.M. and Haubold, H.J.},
  isbn={9781536120424},
  lccn={2017022249},
  series={Mathematics research developments},
  url={https://books.google.com.br/books?id=PbGYswEACAAJ},
  year={2017},
  publisher={Nova Science Publishers, Incorporated}
}

@article{Quevedo2008GeometryThermodynamics,
  author    = {Quevedo, Hernando and V{\'a}zquez, Alejandro},
  title     = {The geometry of thermodynamics},
  journal   = {AIP Conference Proceedings},
  volume    = {977},
  pages     = {165--172},
  year      = {2008},
  doi       = {10.1063/1.2902782},
  url       = {https://doi.org/10.1063/1.2902782}
}

@article{Borges2004,
  author    = {Borges, Ernesto P.},
  title     = {A possible deformed algebra and calculus inspired in nonextensive thermostatistics},
  journal   = {Physica A},
  volume    = {340},
  pages     = {95--101},
  year      = {2004},
  doi       = {10.1016/j.physa.2004.03.082}
}

@article{Campostrini2001,
  author    = {Campostrini, M. and Pelissetto, A. and Rossi, P. and Vicari, E.},
  title     = {Critical behavior of the three-dimensional XY universality class},
  journal   = {Physical Review B},
  volume    = {63},
  pages     = {214503},
  year      = {2001},
  doi       = {10.1103/PhysRevB.63.214503}
}

@article{Osborn2003PRB,
  author  = {Osborn, K. D. and Barber, R. P. and Dynes, R. C.},
  title   = {Critical dynamics of superconducting Bi$_2$Sr$_2$CaCu$_2$O$_{8+\delta}$ films},
  journal = {Physical Review B},
  volume  = {68},
  pages   = {144516},
  year    = {2003},
  doi     = {10.1103/PhysRevB.68.144516}
}

@book{DeGennes1999,
  author    = {De Gennes, P. G.},
  title     = {Superconductivity of Metals and Alloys},
  publisher = {Westview Press},
  address   = {Boulder, CO},
  year      = {1999},
  isbn      = {9780738201016}
}

@article{London1935,
  author    = {London, F. and London, H.},
  title     = {The Electromagnetic Equations of the Supraconductor},
  journal   = {Proceedings of the Royal Society A},
  volume    = {149},
  pages     = {71--88},
  year      = {1935},
  doi       = {10.1098/rspa.1935.0048}
}

@article{Werthamer1966,
  author    = {Werthamer, N. R. and Helfand, E. and Hohenberg, P. C.},
  title     = {Temperature and purity dependence of the superconducting critical field, ${H}_{c2}$. III. Electron spin and spin-orbit effects},
  journal   = {Physical Review},
  volume    = {147},
  pages     = {295--302},
  year      = {1966},
  doi       = {10.1103/PhysRev.147.295}
}

@book{Tarasov2010,
  author    = {Tarasov, Vasily E.},
  title     = {Fractional Dynamics: Applications of Fractional Calculus to Dynamics of Particles, Fields and Media},
  publisher = {Springer},
  address   = {Berlin, Heidelberg},
  year      = {2010},
  isbn      = {978-3-642-14003-7},
  doi       = {10.1007/978-3-642-14003-7},
  url       = {https://doi.org/10.1007/978-3-642-14003-7}
}

@article{Laskin2000FractionalQuantumMechanics,
  author  = {Laskin, Nick},
  title   = {Fractional Quantum Mechanics},
  journal = {Physical Review E},
  volume  = {62},
  number  = {3},
  pages   = {3135--3145},
  year    = {2000},
  doi     = {10.1103/PhysRevE.62.3135}
}

@book{West2020,
  author    = {West, Bruce J.},
  title     = {Fractional Calculus View of Complexity: Tomorrow's Science},
  publisher = {Routledge, Taylor \& Francis Group},
  address   = {Boca Raton, FL},
  year      = {2020},
  isbn      = {9780367737795},
  note      = {Reprint of the 2017 CRC Press edition},
  url       = {https://www.routledge.com/Fractional-Calculus-View-of-Complexity-Tomorrows-Science/West/p/book/9780367737795}
}

@book{erdelyi1981htf,
  author    = {Erdelyi, A.},
  title     = {Higher Transcendental Functions},
  volume    = {3},
  publisher = {R. E. Krieger Publishing Company},
  address   = {Malabar, FL},
  year      = {1981},
  note      = {Bateman Manuscript Project}
}

\end{document}